\documentclass[aps,prd,superscriptaddress]{revtex4}
\usepackage{epsfig,epsf}
\usepackage{amsmath}
\usepackage{amsthm}
\usepackage{amsfonts}
\usepackage{amssymb}
\usepackage{dsfont}
\usepackage{multirow}
\usepackage{appendix}
\usepackage{slashed}
\usepackage[active]{srcltx}
\usepackage{psfrag}
\usepackage{subfigure}
\usepackage[colorlinks,citecolor=blue,urlcolor=red,linkcolor=red]{hyperref}
\usepackage[colorlinks,citecolor=blue,urlcolor=red,linkcolor=red]{hyperref}
\begin{document}
\title{{\Large{\bf Charm meson couplings  in  hard-wall Holographic QCD}}}

\author{\small S. Momeni$^1$  \footnote {e-mail: samira.momeni@ph.iut.ac.ir } and 
\small M. Saghebfar$^2$ \footnote {e-mail: saghebfar@mut-es.ac.ir}}

\affiliation{\emph{$^1$Department of Physics, Isfahan University of
Technology, Isfahan 84156-83111, Iran \\ $^2$Optics-Laser Science and Technology Research Center, Malek Ashtar
University of  Technology, Isfahan , Iran}}

\begin{abstract}
The four- flavor hard- wall holographic QCD is studied to evaluate
the couplings of $(D^{_{-(*-)}}, \bar{D}^{_0}, a_{1}^{_-})$, $(D^{_{-(*-)}}, \bar{D}^{_0}, b_{1}^{_-})$,
$(D_{s}^{_{-(*-)}},\bar{D}^{_{0}}, K_{1A}^{_-})$, $(D_{s}^{_{-(*-)}}, \bar{D}^{_{0}}, K_{1B}^{_{-(*-)}})$,
$(D_{s}^{_{+(*+)}}, {D}^{_+}, K_{1A}^{_0})$, $(D_{s}^{_{+(*+)}}, {D}^{_+}, K_{1B}^{_0})$, $(D^{_{-(*-)}}, \bar{D}^{_{0(*0)}}, \rho^{_-})$,
$(D_{s}^{_{-(*-)}}, \bar{D}^{_{0(*0)}}, K^{_{*-}})$, $({D}^{_{0(*0)}}, \bar{D}^{_{0(*0)}}, \psi)$, $(D_{1}^{_-}, \bar{D}_{1}^{_0}, \pi^{_-})$,
$(D_{s1}^{_-}, \bar{D}_{1}^{_0}, K^{_{-}})$, $({D}_{1}^{_0}, \bar{D}_{1}^{_0}, \eta_{c})$, $(\psi, D^{_{0(*0)}}, D^{_{+}}, \pi^{_-})$,
$(\psi, D^{_{0(*0)}}, \bar{D}^{_0}, \pi^{_0})$, $(\psi, D_{s}^{_{+(*+)}}, D^{_{-}}, K^{_{0}})$, $(\psi, D^{_{0(*0)}}, D^{_{+}}, a_{1}^{_-})$,
$(\psi, D^{_{0(*0)}}, D^{_{+}}, b_{1}^{_-})$, $(\psi, D_{s}^{_{+(*+)}}, D^{_{-}}, K_{1B}^{_{0}})$
 and $(\psi, D_{s}^{_{+(*+)}}, D^{_{-}}, K_{1B}^{_{0}})$  vertices. Moreover, the values of the masses of $D^{_{0(*0)}}$, $D_{s}^{_{-(*-)}}$,
 $\omega$, $\psi$, $D_{1}^{_{0}}$, $D_{1}^{_{-}}$, $K^{0}$, $\eta_{c}$, $D_{s1}^{_{-}}$ and $\chi_{_{c1}}$ as well as the decay constant
 of $\pi^{-}$, $D^{_{-(*-)}}$, $K^{-}$, $\rho^{-}$, $D_{1}^{_{-}}$ , $a_{1}^{-}$  and $D_{s}^{_{-(*-)}}$
 are estimated in this study. A comparison is also made between our results and the experimental values of the
  masses and decay constants. Our results for  strong couplings are also compared with the
  3PSR and  LCSR predictions.
\end{abstract}


\maketitle

\section{Introduction}
In recent investigations,  the strong interaction of charmed hadrons
 among themselves and with other particles have received remarkable attention.
In phenomenology  of the high energy physics,
charm meson vertices play a perfect role in
 meson interactions.

The charmed meson vertices help us to investigate
the final-state interactions in  hadronic B decays. In these studies,  charm
mesons are considered as the intermediate states which   lead the
long distance  effect on the values of the branching ratios for
non-leptonic $B$ meson decays, are studied
in \cite{Isola1, Isola2, Khosravi2018, Janbazi2018, Colangelo2004,
Colangelo20042, Ladisa2004, Cheng2005, Deandrea2006}. On the other hand, the strong couplings between
charm mesons and other hadrons, can help us to
study the  production of  $J/\psi, \psi(2s), \cdots$, in heavy ion collisions and
absorption of these states
in hadronic matter such as   nucleons and  light mesons \cite{Bracco2005, Holanda2007}. Now a days, different
theoretical methods are used to consider vertices involving charmed mesons.
 $D^{*}\,D\,\pi$, $D^{*}\,D\,\gamma$, $D\,D\, \rho$, $D^{*}\,D^{*}\,\rho$  vertices are analyzed
 via lattice QCD approach  in \cite{Can2013, Abada2002, Becirevic2011, Becirevic2013}.
Moreover, $D^* D^* \rho$ \cite{MEBracco}, $D^* D \pi$ \cite{FSNavarra,MNielsen}, $D
D \rho$ \cite{MChiapparini}, $D^* D \rho$ \cite{Rodrigues3}, $D D
J/\psi$ \cite{RDMatheus},  $D^* D J/\psi$ \cite{RRdaSilva},
$D^*D_sK$, $D^*_sD K$, $D^*_0 D_s K$, $D^*_{s0} D K$ \cite{SLWang},
$D^*D^* P$, $D^*D V$, $D D V$ \cite{ZGWang}, $D^* D^* \pi$
\cite{FCarvalho}, $D_s D^* K$, $D_s^* D K$ \cite{ALozea}, $D D
\omega$ \cite{LBHolanda},  $D_s D_s V$, $D^{*}_s D^{*}_s V$
\cite{KJ,KJ2},  $D_1D^*\pi, D_1D_0\pi,  D_1D_1\pi$ \cite{Janbazi} and  $D D A, D^{*} D A$ \cite{momeni2020},  vertices are often
studied  via the three point sum rule (3PSR) and the light cone QCD sum rule (LCSR) methods.

In recent years, a relatively new
approach  named the anti-de Sitter space/quantum chromodynamics (AdS/QCD) correspondence has been utilized
to predict the form factors and couplings for the  hadronic systems. This method is inspired
from correspondence between a type IIB string theory
and  super Yang-Mills theory in the large
$N_{c}$ limit with  $\mathcal{N}=4$ \cite{Maldacena1998, Witten1998, Gubser1998}.
In this approach,
corresponding  to every field in the $\rm{AdS}_{5}$ space, an
operator is defined in 4 dimensional gauge theory,  and
the correlation functions involving $n$ currents are related
 to the $5$D action by functional differentiation  with respect to their
$n$ sources \cite{Witten1998, Gubser1998, Grigoryan2007, Abidin2008}.
 Utilizing (AdS/QCD) correspondence approach interesting results are
 reported as the  masses, couplings,
electromagnetic and gravitational form factors of mesons
 \cite{Polchinski2002, Polchinski20022, Brodsky2004, Teramond2005,
 Brodsky2006, Brodsky2008, Grigoryan20072, Grigoryan20073, Grigoryan2008, Kwee2008,Kwee20082, Boschi-Filho2006, Abidin20082, Abidin20091}.
This method is also utilized to predict $K_{\ell 3}$  transition form factors in \cite{Abidin2009}.
In addition,  the strong couplings $g_{\rho^{n} \rho \rho}$, $g_{\rho^{n} K K}$, $g_{\rho^{n}K^{*}K^{*}}$,
$g_{\rho^{n} D D}$ and $g_{\rho^{n}D^{*}D^{*}}$ are analyzed in a hard wall holographic QCD in \cite{Bayona2017}.

Our goal in this paper is to extract the couplings of $(D, D, A)$, $(D^{*}, D, A)$, $(D, D,V)$, $(D^{*}, D, V)$
$(D^{*}, D^{*}, V)$, $(D_{1}, D_{1}, P)$, $(\psi, D,  D, A)$, $(\psi, D^{*}, D, A)$, $(\psi, D, D, P)$
and $(\psi,  D^{*},  D,  P)$ in  hard wall holographic QCD with four flavors.
The paper is organized as follows: In Sec. \ref{sec.1}, our model including pseudoscalar, vector and axial vector mesons
is introduced. In Sec. \ref{sec.2}, the wavefunctions and the decay constant of studied mesons are extracted from
our model. The strong couplings for three and four- meson  vertices  derived in  Sec. \ref{sec.3anf4} and Sec. \ref{sec.num}
is reserved for numerical analysis. Our prediction for masses, decay constants, wavefunctions and the strong couplings
are presented in this section. For a better analysis, a comparison is made between
our estimations and the results of other methods.

\section{The AdS/QCD model involving pseudoscalar, vector and axial vector mesons}\label{sec.1}
In this section we introduced our model in $5$ dimensions involving pseudoscalar, vector and axial vector mesons.
In this paper, the metric of 5 dimensional Anti-de Sitter space is chosen in Poincare coordinates as:
\begin{eqnarray}\label{eq.1}
ds^2=\frac{1}{z^2}\left(-dz^2+\eta_{\mu\nu}dx^{\mu}\,dx^{\nu}\right),
\end{eqnarray}
where $ \mu,\nu=0,1,2,3$. Moreover, $\eta_{\mu\nu}=\rm{diag} (1,-1,-1,-1)$ is the usual Minkowski metric in 4 dimensions.
In hard-wall model,  the radial coordinate $z$ varies in the range    $ (\varepsilon, z_{0})$, where
the lower bound  $z=\varepsilon $ (with $\varepsilon\to 0$) gives the asymptotic feature of QCD  and the
IR cut-off $z_{0}\approx 1/\Lambda_{\rm{QCD}}$ is used to simulate QCD confinement.

We will consider the 5D action  proposed in Ref \cite{Erlich2005}. In this
action the  $N_{f}$ gauge fields $L^{\mu, a}$, $R^{\mu, a}$   and a scalar field $X$ correspond to
5D fields for current operators $J^{\mu, a}_{L/R}=\bar{q}_{L/R}\,\gamma^{\mu}\,t^{a}\,q_{L/R}$ and
 $\bar{q}_{L}\,q_{R}$  from 4D theory, respectively. In $J^{\mu, a}_{L/R}$ definition, $q$ is quark field and
   $q_{L/R}=(1\pm\gamma_{5})q$ are the left handed (L) and the right handed (R) quarks. Moreover,
   $t^{a}$ (with $a=1,\cdots N_{f}^{2}-1$) are the generators of the
  SU$(N_{f})$ group which are related to the Gell-Mann matrices $\lambda^{a}$ by
  $\lambda^{a}=2\,t^{a}$.
 In this paper, we take $N_{f}=4$ and the 5D action with
 SU(4)$_L \otimes$\,SU(4)$_R$ symmetry can be written as
\begin{eqnarray}\label{eq.2}
S = \int d^5 x \, \sqrt{g} \, {\rm Tr}\Big\{(D_{M}X)\,(D^{M}X)^{\dag} +3\left|X\right|^2-\frac{1}{4g_5^2}\left(
L^{MN}\,L_{MN}+R^{MN}\,R_{MN}\right)\Big\}  .
\end{eqnarray}
where $D_{M} X=\partial_M X -iL_M X +i X R_M$ is the
 covariant derivative of the scalar  field $X$. In addition,  the  strength of the
non-Abelian $L$ and $R$ fields are defined as
\begin{eqnarray}\label{eq.3}
L_{MN} &=& \partial_M L_N - \partial_N L_M - i \left [ L_M , L_N \right ], \nonumber \\
R_{MN} &=& \partial_M R_N - \partial_N R_M - i \left [ R_M , R_N \right ],
\end{eqnarray}
with $L_{M}=L_{M}^{a}\,t^{a} $ and $R_{M}=R_{M}^{a}\,t^{a}$.
The  left and right hand gauge fields  can also be written in terms of
the vector (V) and the axial vector field $A$ , in the form $L=V+A$ and $R=V-A$.
The scalar field  X can be expanded  as
\begin{eqnarray}\label{eq.3}
X= e^{i \pi^a t^a}\,  X_0 \,e^{i \pi^a t^a}
\end{eqnarray}
where $X_0$ is the classical part and $\pi$ contains the
fluctuations. With flavor symmetry, $X_{0}$ is a multiple of the unit matrix and $X = e^{2i \pi^a t^a}  X_0$ can be obtained.
This choice for the scalar field is used in
 \cite{Shock2006} with $N_{f}=2$,  and
flavor symmetry is assumed to estimate
 masses and decay constants for the light and strange mesons.  Their  model  predicts
 good results for the more excited strange mesons observables.
In \cite{Katz2007} the part
of the action that mixes the axial vectors with the pseudoscalars is just considered and the  U$(1)$ problem is studied.
All  parameters in the mentioned model can be determined by the experimental
masses of the $\pi^{0}$, $K^{0}$ and $\rho$ mesons,
and the pion decay constant $f_{\pi}$.

In general , using equation of motions and turning off all fields except $X_{0}(z)$, one obtains
\begin{eqnarray}\label{eq.4}
2 X_0(z) = \zeta M\, z + \frac{\Sigma}{\zeta}\, z^3,
\end{eqnarray}
where $M$ and $\Sigma$ are  the quark-mass  and the
quark condensates $\left<\bar q q\right>$ matrices, respectively. For $N_{f}=4$ we take $M={\rm diag}(m_u , m_d , m_s , m_c)$
and  $\Sigma={\rm diag}(\sigma_u ,  \sigma_d , \sigma_s , \sigma_c)$.
Moreover in Eq. (\ref{eq.4}), $\zeta=\sqrt{N_c}/2\pi$ is  the normalization
parameter introduced in Ref. \cite{Cherman2009}. Note that for the light-quark sectors in the SU$(2)$ isospin
symmetry,    $m_{d} = m_{u} $ and $\sigma_u=\sigma_d$ are assumed in \cite{Abidin2009,Bayona2017}.
Eq. (\ref{eq.4}) is used  in Refs. \cite{Colangelo2008, Abidin2009, Maru2009,  Huseynova2019}
and in this paper we shall use it.

\section{Wave functions, masses and the decay constants for the pseudoscalar, vector and axial vector mesons }\label{sec.2}
Expanding the action in Eq. (\ref{eq.2}) up to second  order in
  vector (V), axial vector (A) and  pseudoscalar field $(\pi)$, we obtain
\begin{eqnarray} \label{eq.5}
S&=&\int\, d^5 x  \bigg\{\sum_{a=1}^{15} \frac{-1}{4g_5^2 z} (\partial_M V^a_N-\partial_N V^a_M)\,
(\partial^M V^{N}_{a}-\partial^{N} V^{M}_{a})+\frac{ {M_V^a}^2}{2z^3}{V^a_M}\,{V^M_a}\nonumber\\
&&\frac{-1}{4g_5^2 z} (\partial_M A^a_N-\partial_N A^a_M)\,(\partial^M A^{N}_{a}-\partial^{N}
A^{M}_{a})+\frac{{M_A^a}^2}{2z^3}(\partial_M\pi^a-A^a_M)\,(\partial^M\pi_{b}-A^M_{b}) \bigg\},
\end{eqnarray}
where we have defined:
\begin{eqnarray}\label{eq.defmva}
{M_V^a}^2 \delta^{ab} &=&-2\, {\rm Tr} \left.( [t^a , X_0] [ t^b, X_0] \right.),\nonumber\\
{M_A^a}^2 \delta^{ab} &=&2\, {\rm Tr} \left.( \{ t^a , X_0 \} \{ t^b , X_0 \} \right.).
\end{eqnarray}
Using
\begin{eqnarray}
v_q (z) = \zeta m_q z + \frac{1}{\zeta} \sigma_q z^3 \, \,  , \, \,  q=(u, d, s, c),
\end{eqnarray}
 the   values  reported in Table. \ref{mav}, are obtained for ${M_V^a}^2$ and ${M_A^a}^2$.
\begin{table}[!th]
\caption{The values of  ${M_V^a}^2$ and ${M_A^a}^2$ with
$v_q (z) = \zeta m_q z + \frac{1}{\zeta} \sigma_q z^3$ for    $q=(u, d, s, c)$} \label{mav}
\begin{tabular}{|c||c|c||c||c|c||c||c|c|}
\hline
&&&&&&&&\\
$a$ &${M_V^a}^2$&${M_A^a}^2$&$a$&${M_V^a}^2$&${M_A^a}^2$&$a$&${M_V^a}^2$&${M_A^a}^2$\\
&&&&&&&&\\
\hline
&&&&&&&&\\
$(1, 2)$ &$\frac{1}{4} (v_u - v_d)^2$&$\frac{1}{4} (v_u + v_d)^2$&$(6, 7)$&$\frac{1}{4} (v_d - v_s)^2$&$\frac{1}{4} (v_d + v_s)^2$&
$(11, 12)$&$\frac{1}{4} (v_d- v_c)^2$&$\frac{1}{4} (v_d+ v_c)^2$\\
&&&&&&&&\\
\hline
&&&&&&&&\\
$3$ &$ 0$&$\frac{1}{2} (v_u^2 + v_d^2)$&$8$&$0$&
$\frac{1}{6} (v_u^2+v_d^2+4 v_s)^2 $&$(13, 14)$&$\frac{1}{4} (v_c - v_s)^2$&$\frac{1}{4} (v_c + v_s)^2$\\
&&&&&&&&\\
\hline
&&&&&&&&\\
$(4, 5)$ &$\frac{1}{4} (v_u - v_s)^2 $&$\frac{1}{4} (v_u + v_s)^2$&$(9, 10)$&
$\frac{1}{4} (v_u - v_c)^2 $&$\frac{1}{4} (v_u + v_c)^2$&
$15$&$0$&$\frac{1}{12} ( v_u^2 +v_d^2+ v_s^2 + 9 v_c^2 )$\\
&&&&&&&&\\
\hline
\end{tabular}
\end{table}

Now we are ready to derive equation of the motion for the  vector, axial vector and  pseudoscalar fields.
\subsection{Wave functions  }
In this subsection we study wave functions of  vector,   axial vector and pseudoscalar mesons. We start with the vector field,
which  satisfies the following equation of motion
\begin{eqnarray}\label{eq.eomv1}
\eta^{ML}\partial_M\left(\frac{1}{z}\left(\partial_L V^a_N-\partial_N V^a_L\right)\right)+\frac{\alpha^a(z)}{z}V^a_N=0.
\end{eqnarray}
Where $z^{2}\alpha^a(z) = g_5^2 {M_V^a}^2$.
For the transverse part, choosing $\partial^\mu V^a_{\mu\perp}(x,z)=0$,
the following result is obtained:
\begin{eqnarray}\label{eq.eomv2}
\left(\partial_z\frac{1}{z}\partial_z+\frac{q^2-\alpha^a}{z}\right) V^a_{\mu\perp}(q,z)=0,
\end{eqnarray}
 Here,  $q$ is the Fourier variable conjugate to the 4 dimensional
coordinates, $x$.
The transverse part of the vector
field can be written as $V^a_{\mu\perp}(q,z)= V^{0a}_{\mu\perp}(q) {\cal V}^a(q^2,z)$ where
$V^{0a}_{\mu\perp}$  and ${\cal V}^a$ are boundary
values at UV  and bulk-to- boundary
propagator, respectively.  ${\cal V}^a(q^2, z)$ satisfies the same
equation as $V^{a}_{\mu\perp}(q, z)$ with the boundary conditions
${\cal V}^a(q^2, \varepsilon) = 1$ and $\partial_z {\cal V}^a(q^2, z_{0}) = 0 $.

The longitudinal parts of the vector field,  defined as $V^a_{\mu\parallel}=\partial_\mu \xi^a$,
and $V^a_z=-\partial_z \tilde \pi^a$,  are  coupled as follows:
\begin{eqnarray}
&&-q^2\partial_z \tilde \phi^a(q^2,z) +\alpha^a\partial_z \tilde \pi^a(q^2,z)=0\,,\label{eq.eomv3}\\
&&\partial_z \bigg(\frac{1}{z}\partial_z\tilde \phi^a(q^2,z) \bigg)
 -\frac{\alpha^a}{z}\big(\tilde \phi^a(q^2,z)-\tilde \pi^a(q^2,z)\big)=0,\label{eq.eomv4}
\end{eqnarray}
where the boundary conditions
are $\tilde \phi^a(q,\varepsilon)=0$, $\tilde \pi^a(q,\varepsilon)=-1$ and
 $\partial_z \tilde\phi^a(q^2,z_0)=\partial_z\tilde\pi^a(q^2,z_0)=0$.

 In general form of differential
equations Eqs. (\ref{eq.eomv2}, \ref{eq.eomv3}),
 $\mathcal{V}^{a}(q^2, z)$, $\tilde \phi^a(q^2, z)$ and $\tilde \pi^a(q^2, z)$ can be solved numerically.
We expect  that, normalizable
modes of Eq. (\ref{eq.eomv2})  describe the vector mesons while,
 Eqs. (\ref{eq.eomv3}) and  (\ref{eq.eomv4}) are utilized to study the scalar ones.
In this, paper the scalar mesons are not considered.

To obtain the  wave functions of the axial vector and pseudoscalar mesons,
 the variation over the axial vector field ($A_{M}^{a}$) of Eq. (\ref{eq.5}),  is taken.
The transverse part of the axial vector field satisfies the following equation of motion:
\begin{eqnarray}\label{eq.eom}
\left(\partial_z\frac{1}{z}
\partial_z+\frac{q^2-\beta^a}{z}\right) A^a_{\mu\perp}(q,z)=0,
\end{eqnarray}
where $z^{2}\beta^a(z) = g_5^2 \,{M_A^a}^2$. Moreover, the  gauge choices
$\partial^\mu A^a_{\mu\perp}(x,z)=0$ and $A_{z}^{a}=0$ are  imposed
in the Fourier transform. Note that $A_{\mu}^{a}=A_{\mu\perp}^{a}+\partial_\mu \phi^{a}$  is used to separate
 the
transverse and
longitudinal parts of the axial vector field.

The transverse part $A_{\mu\perp}^{a}$, can be written as
$A^a_{\mu\perp}(q,z)={{A}_{\mu\perp}^{a0}(q)} \mathcal A^{a}(q^2,z)$.
To obtain $\mathcal A(q^2, z)$, we set $\mathcal A^{a}(q^2, \varepsilon)=1$ for the UV boundary
and for the IR boundary we choose Neumann boundary condition $\mathcal A^{a}(q^2, z_{0})=0$.
This part describes the axial vector states.

The longitudinal part of the
axial-vector field $\phi^{a}$  and the $\pi^{a}$ describe the pseudoscalar fields and
 satisfy the following
equations
\begin{eqnarray}	
&&-q^2\partial_z \phi^a(q^2,z)+\beta^a(z)\partial_z \pi^a(q^2,z)=0\,,\label{longtueq1}\\			
&&\partial_z \left(\frac{1}{z}\partial_z\phi^a(q^2,z) \right)  -
\frac{\beta^a(z)}{z}\left(\phi^a(q^2,z)-\pi^a(q^2,z)\right)=0\,,\label{longtueq2} 					
\end{eqnarray}
where the  boundary
conditions are $\phi^a(q^2,\varepsilon)=0$, $\pi^a(q^2,\varepsilon)=-1$, and
$\partial_z\phi^a(q^2,z_0)=\partial_z\pi^a(q^2,z_0)=0$.

We  finish this subsection by  writing the SU(4) vector $V$, axial vector $A$ and pseudoscalar $\pi$
 meson matrices  terms of the
charged states as:
\begin{widetext}
\begin{eqnarray*}
 V &=&  V^a t^a  = \frac{1}{\sqrt 2}
\left ( \begin{matrix}
  \frac{\rho^0}{\sqrt{2}} + \frac{\omega'}{\sqrt{6}} + \frac{\psi}{\sqrt{12}}  &  \rho^{_{+}}  &  K^{_{*+}}  &  \bar D^{_{*0}}  \\
  \rho^{_{-}}   & -\frac{\rho^0}{\sqrt{2}}  + \frac{\omega'}{\sqrt{6}}  + \frac{\psi}{\sqrt{12}}   &  K^{_{*0}}  &  D^{_{*-}}  \\
   K^{_{*-}}  &  \bar K^{_{*0}}  & - \sqrt{\frac23} \omega' + \frac{\psi}{\sqrt{12}}  &  D_s^{_{*-}}  \\
   D^{_{*0}} &  D^{_{*+}}  &  D_s^{_{*+}}  & - \frac{3}{\sqrt{12}} \psi
 \end{matrix} \right ), \\
A &=&  A^a t^a  =  \frac{1}{\sqrt 2} \left(\begin{array}{cccc}
 \frac{a^0_1+b^0_1}{\sqrt{2}}  + \frac{f_1+f_{1}'}{\sqrt{6}}+\frac{\chi_{c1}}{\sqrt{12}}
 & a^{_{+}}_1+ b^{_{+}}_1 & K_{1A}^{_{+}}+K_{1B}^{_{+}}& \bar {D}_{1}^{_{0}} \\
a_1^{_{-}}+b_1^{_{-}}&- \frac{ a^{_{0}}_1+b^{_{0}}_1}{\sqrt{2}} +\frac{f_1+f_{1}'}{\sqrt{6}}+\frac{\chi_{c1}}{\sqrt{12}}
& K_{1A}^{_{0}}+K_{1B}^{_{0}}& D_{1}^{_{-}}\\
K_{1A}^{_{-}}+K_{1B}^{_{-}}& \bar{K}_{1A}^{_{0}}+\bar{K}_{1B}^{_{0}}  & -\sqrt{\frac23}(f_1+f_{1}')+\frac{\chi_{c1}}{\sqrt{12}}& D_{s1}^{_{-}}\\
{D}_{1}^{_{0}}&D_{1}^{_{+}}&D_{s1}^{_{+}}&- \frac{3}{\sqrt{12}}\chi_{c1}
\end{array}
\right), \\
\pi &=& \pi^a T^a = \frac{1}{\sqrt 2}
\left (\begin{matrix}
  \frac{\pi^{_{0}}}{\sqrt{2}}  + \frac{\eta}{\sqrt{6}}  + \frac{\eta_c}{\sqrt{12}}  &  \pi^{_{+}}  &  K^{_{+}}  &  \bar D^{_{0}}  \\
  \pi^{_{-}}   & -\frac{\pi^0}{\sqrt{2}}  + \frac{\eta}{\sqrt{6}}  + \frac{\eta_c}{\sqrt{12}}   &  K^{_{0}}  &  D^{_{-}} \\
   K^{_{-}}  &  \bar K^{_{0}}  & - \sqrt{\frac23} \eta + \frac{\eta_c}{\sqrt{12}}  &  D_s^{_{-}}  \\
   D^{_{0}} &  D^{_{+}}  &  D_s^{_{+}}  & - \frac{3}{\sqrt{12}} \eta_c
 \end{matrix} \right ).
\end{eqnarray*}
\end{widetext}
It should be noted that $K_{1A}$ and $K_{1B}$ are not physical states.
 The physical states of  $K_{1}(1270)$ and $K_{1}(1400)$  mesons are
related to these states in terms of a mixing angle $\theta_{K}$  as follows:
\begin{eqnarray}\label{eql1mix}
K_1(1270) &=&\sin\theta_K \,K_{1A} +
\cos\theta_K\, K_{1B},\nonumber\\
K_1(1400)&=&\cos\theta_K \,K_{1A} - \sin\theta_K\,
K_{1B}.
\end{eqnarray}
The
mixing angle $\theta_K$  can be determined by the experimental data.
There are various approaches to estimate the mixing angle. The
result $35^\circ < |\theta_K| < 55^\circ$ was found in Ref.
\cite{Burakovsky}, while  two possible solutions with
$|\theta_K|\approx 33^\circ$ and $57^\circ$ were obtained in Ref.
\cite{Suzuki}.

\subsection{Decay constants}
To evaluate the  decay constant of the vector mesons in the above mentioned model, the
two- point functions are needed. According to AdS/QCD correspondence, two-point functions can be calculated
 by evaluating the action, Eq. (\ref{eq.5})
with the classical solution and taking the functional
derivative over $V^0_\mu$ twice as:
\begin{eqnarray}\label{eq.twopoint}
\langle 0 | {\mathcal T}
\{J^{a\mu}_{V\perp}(x)\,J^{b\nu}_{V\perp}(y)\}
| 0 \rangle
= -i
\frac{  \delta^2 \rm{S}(VV)}
{\delta V^{a0}_{\mu\perp} (x)  \delta V^{b0}_{\nu\perp}(y)}	,
\end{eqnarray}
In the LHS of Eq. (\ref{eq.twopoint}), we insert one complete set
intermediate states with the same quantum numbers as the meson currents,
 and use the vector mesons decay constants definition as:
\begin{eqnarray}\label{elementsdef3}
\langle0|J_{V\perp}^{\nu a}|V^{a'}(p, \varepsilon)\rangle&=&f_{V}\,\varepsilon^{\nu}\,\delta^{aa'},
\end{eqnarray}
where $f_{V}$ and $\varepsilon$ are the decay constant and the
  polarization vector for vector meson $V(p, \varepsilon)$, respectively.
 After performing the Fourier transformation
\begin{eqnarray}\label{eq.twopoint4}
i\,\int d^{4}x\,e^{ip x}\,{\langle 0 | {\mathcal T}
\{J^{a\mu}_{V\perp}(x)\,J^{b\nu}_{V\perp}(0)\}
| 0 \rangle}=\sum_{n}\frac{f_{V^{n}}^2\,\delta^{ab}}{p^2-m_{V^{n}}^2}\Pi^{\mu\nu}
\end{eqnarray}
is obtained.
Where $\Pi^{\mu\nu}=\left(\eta^{\mu\nu}-p^\mu p^\nu/p^2\right)$ is transverse projector.
In the RHS of Eq. (\ref{eq.twopoint}),
 $\rm{S}(VV)$ contains two vector mesons and can be obtained by inserting the solution for
$V^{a}_{M}$ back into the action. After applying Fourier transformation, in the final result,
only the contribution of the surface term  at $z = \epsilon$  remains as:
\begin{eqnarray}\label{eq.svv}
S (VV) = \int \frac{d^4 p}{(2\pi)^4}
V^{c0}_{\lambda\perp}(p)V^{c0\lambda}_{\perp}(p)\left(-\frac{\partial_z V(p, z)}{2g_5^2z} \right)_{z=\epsilon}.
\end{eqnarray}

On the other hand,  using Green's function formalism
to solve Eq. (\ref{eq.eomv2}), the bulk-to-boundary propagator can be
written
as a sum over
vector mesons poles:
\begin{eqnarray}\label{eq.exv}
{\cal V}^a(q^2,z)=\sum_n \frac{-g_5 f_{_{V^{n}}} \psi^a_{_{V^{n}}}(z)}{q^2-m_{_{V^{n}}}^2}\,,
\end{eqnarray}
where boundary conditions for the $n^{{th}}$ vector meson's wave function
are $\psi_{_{V^{n}}}(\epsilon)=0$ and $\partial_{z}\psi_{_{V^{n}}}(z_{0})=0$. Moreover
the normalization condition  is $\int (dz/z) {\psi^a_{_{V^{n}}}}^2=1$.
Using Eqs. (\ref{eq.twopoint}-\ref{eq.exv}),  the decay constant of the $n^{{th}}$ mode of the vector
meson  is obtained as:
\begin{eqnarray}\label{eq.vde}
f_{V^{n}}=\frac{\partial_{z}\psi_{_{V^{n}}}}{g_5\,z}\bigg|_{z=\epsilon}.
\end{eqnarray}

For the axial vector and the pseudoscalar states, the decay constants are defined as:
\begin{eqnarray}
\langle0|J_{V\perp}^{\nu b}|A^{b'}(p, \varepsilon')\rangle&=&f_{A}\,\varepsilon'^{\nu}\,\delta^{bb'},\label{decayaxial}\\
\langle0|J_{A\parallel}^{\nu d}|\phi^{d'}(p)\rangle&=&i f^d\, p_\nu \delta^{dd'}\label{decayp}.
\end{eqnarray}
To evaluate the decay constants of the
vector mesons  and the pseudoscalar ones, the following Green's functions are used:
\begin{eqnarray}
{\cal A}^a(q^2, z)&=&\sum_n \frac{-g_5 f_{_{A^{n}}} \psi^a_{_{A^{n}}}(z)}{q^2-m_{_{A^{n}}}^2}\,,\nonumber \\
\phi^a(q^2, z)&=&\sum_n\frac{- g_5 {m^a_n}^2 f^a_n\phi^a_n(z)}{q^2-{m^a_n}^2}\,,
\nonumber \\
\pi^a(q^2, z)&=&\sum_n\frac{- g_5 {m^a_n}^2 f^a_n\pi^a_n(z)}{q^2-{m^a_n}^2}\,,	
\end{eqnarray}
where for the $\psi^a_{_{A^{n}}}(z)$ the boundary conditions are similar to  $\psi^a_{_{V^{n}}}(z)$. For the
pseudoscalar meson's wave functions,   $\phi^a_{n}(\varepsilon)=\pi^a_{n}(\varepsilon)=0$ and
$\partial_z\phi^a_{n}(z_0)=\partial_z\pi^a_{n}(z_0)=0$ are the boundary conditions. The similar method
is used to calculate the vector mesons decay constants, the following results
can be obtained for the axial vector mesons and the pseudoscalar states decay constant, respectively:
\begin{eqnarray}\label{eq.aandpide}
f_{A^{n}}&=&\frac{\partial_{z}\psi_{_{A^{n}}}}{g_5\,z}\bigg|_{z=\epsilon},\\
f^a_n&=&-\frac{\partial_{z}\phi^a_n}{g_5\,z}\bigg|_{z=\epsilon},
\end{eqnarray}

\section{Strong coupling constants from three and four point functions}\label{sec.3anf4}
\begin{figure}[!th]
\includegraphics[width=4.5cm,height=4cm]{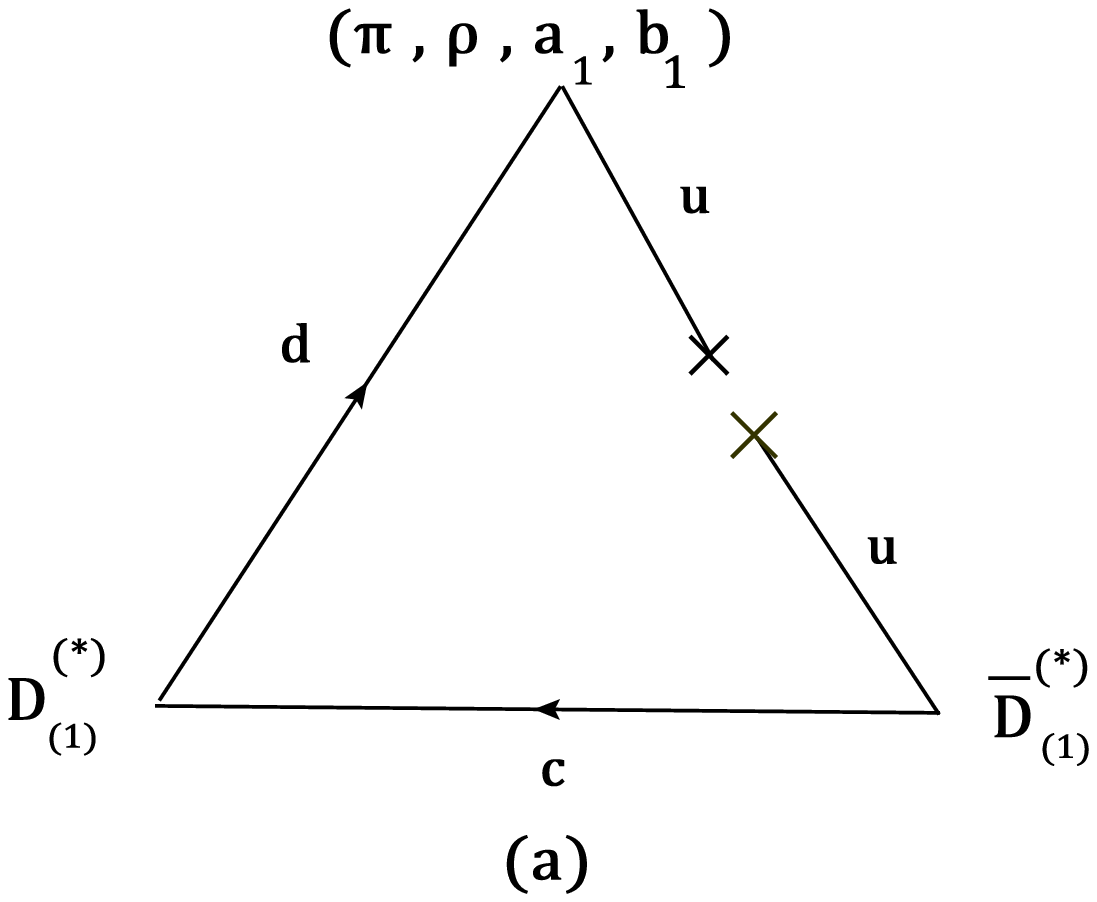}
\includegraphics[width=4.5cm,height=4cm]{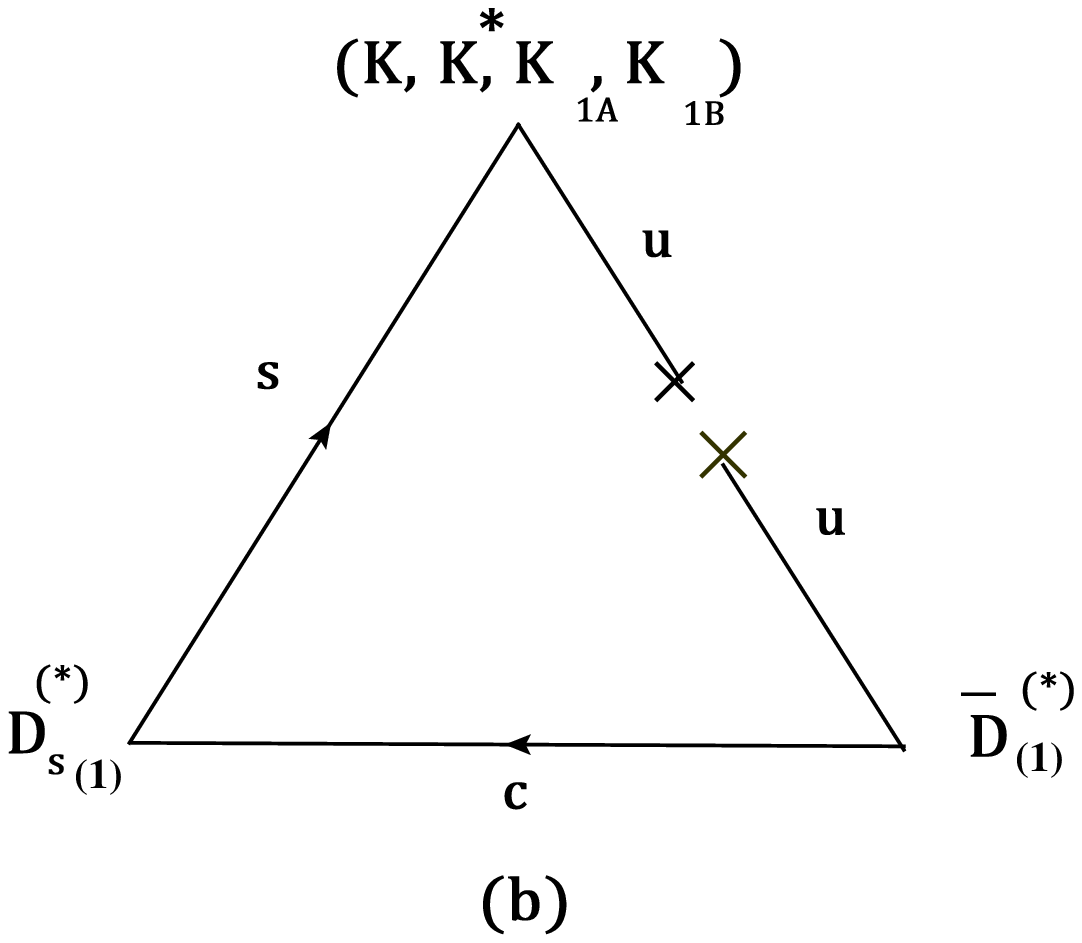}
\includegraphics[width=4.5cm,height=4.3cm]{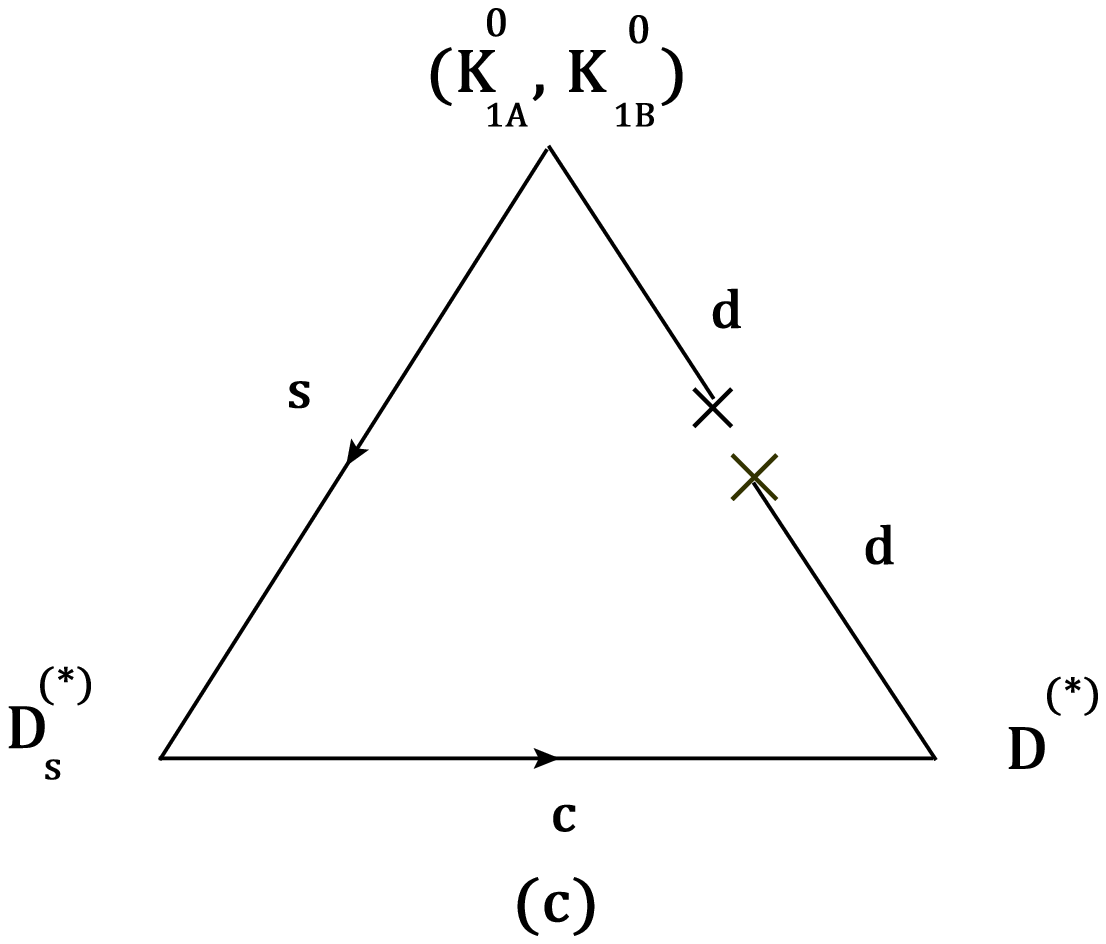}
\caption{3-particle diagrams show $(D^{(*)}, D, A)$, $(D^{(*)}, D^{(*)}, V)$ and $(D_{1}, D_{1}, P)$
vertices. } \label{3pv}
\end{figure}

In this section, we study the  triplet  and quadratic  vertices including
charm, vector, axial vector and  pseudoscalar mesons.
The corresponding diagrams for triplet vertices are given in
Fig. \ref{3pv}. The vertices   $(D^{_{-}}, \bar{D}^{_0}, a_{1}^{_-})$,
$(D^{_-}, \bar{D}^{_0}, b_{1}^{_-})$, $(D^{_{*-}}, \bar{D}^{_0}, a_{1}^{_-})$, $(D^{_{*-}}, \bar{D}^{_0}, b_{1}^{_-})$,
$(D^{_-}, \bar{D}^{_0}\,\rho^{_-})$, $(D^{_{*-}}, \bar{D}^{_0}\,\rho^{_-})$, $(D^{_{*-}}, \bar{D}^{_{*0}}, \rho^{_-})$ and
 $(D_{1}^{_-}, \bar{D}_{1}^{_0}, \pi^{_-})$  can be describe with  diagram (a) while diagram (b)
 is used to explain $(D_{s}^{_-}, \bar{D}^{_{0}}, K_{1A}^{_-})$, $(D_{s}^{_-}, \bar{D}^{_0}, K_{1B}^{_-})$,
$(D_{s}^{_-}, \bar{D}^{_0}, K^{_{*-}})$,
$(D_{s}^{_{*-}}, \bar{D}^{_0}, K_{1A}^{_-})$, $(D_{s}^{_{*-}}, \bar{D}^{_0}, K_{1B}^{_{-}})$,
$(D_{s}^{_{*-}}, \bar{D}^{_0}, K^{_{*-}})$,  $(D_{s}^{_{*-}}, \bar{D}^{_{*0}}, K^{_{*-}})$,
 $(D_{s}^{_{*-}}, \bar{D}^{_{0}}, K^{_{*-}})$ and $(D_{s}^{_{*-}}, \bar{D}^{_{*0}}, K^{_{*-}})$ vertices. Finally,  diagram (c)
shows $(D_{s}^{_+}, {D}^{_+}, K_{1A}^{_0})$, $(D_{s}^{_+}, {D}^{_+}, K_{1B}^{_0})$,
 $(D_{s}^{_{*+}}, {D}^{_+}, K_{1A}^{_0})$ and  $(D_{s}^{_{*+}}, {D}^{_+}, K_{1B}^{_0})$  vertices.

\begin{figure}[!th]
\includegraphics[width=4.5cm,height=4cm]{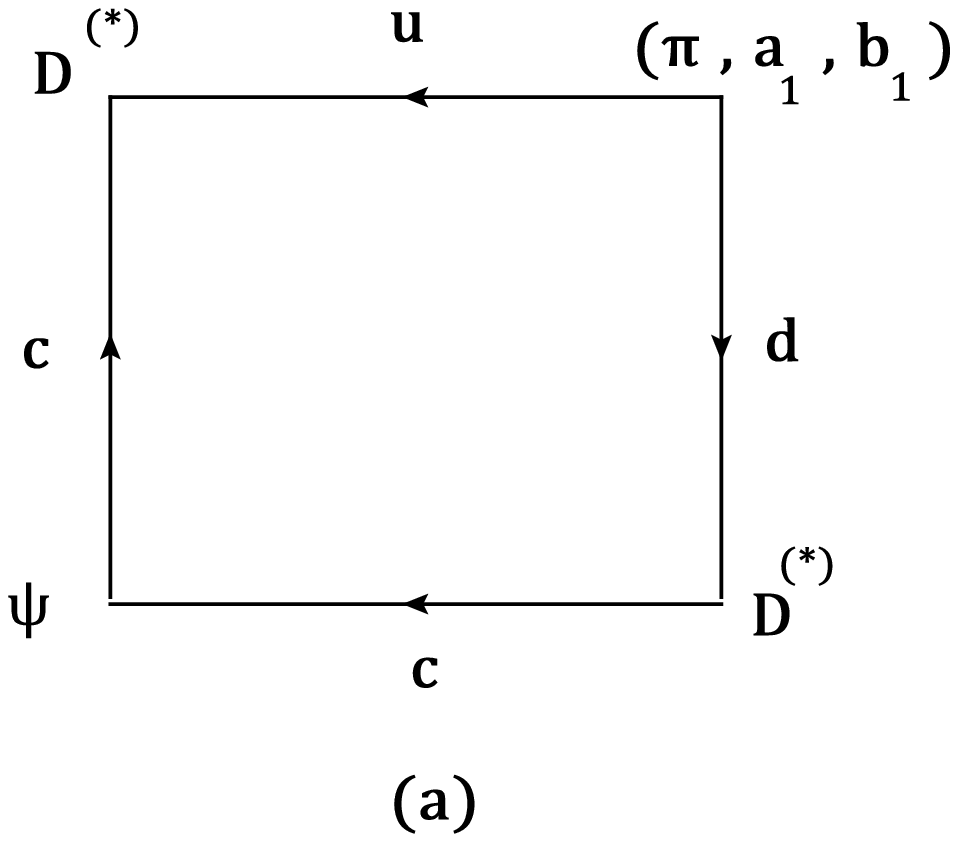}
\includegraphics[width=4.5cm,height=4cm]{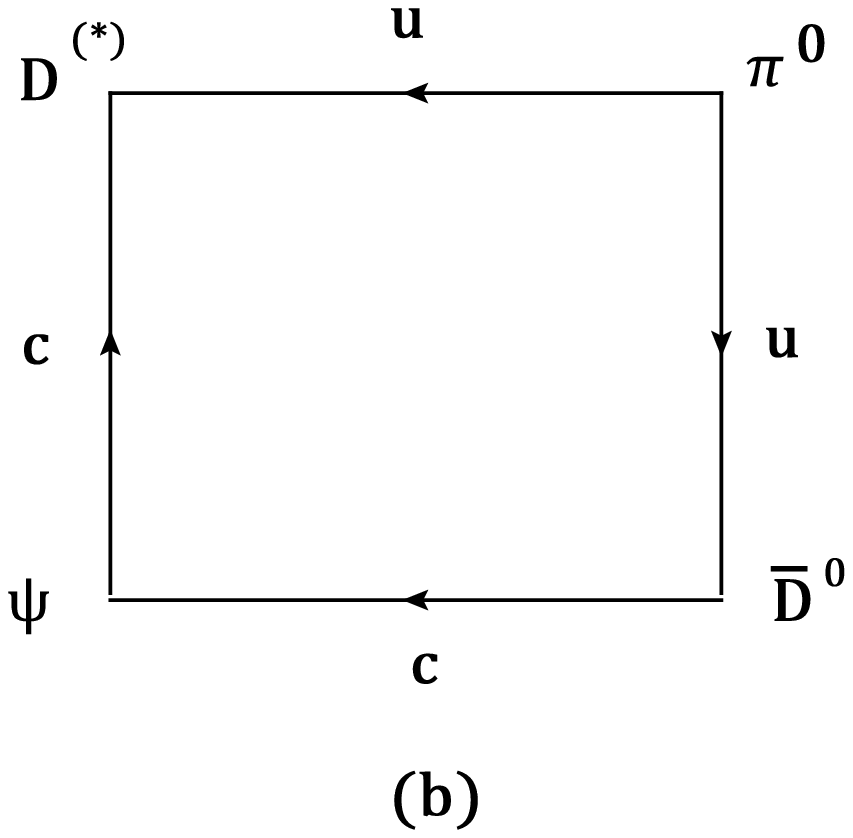}
\includegraphics[width=4.5cm,height=4.2cm]{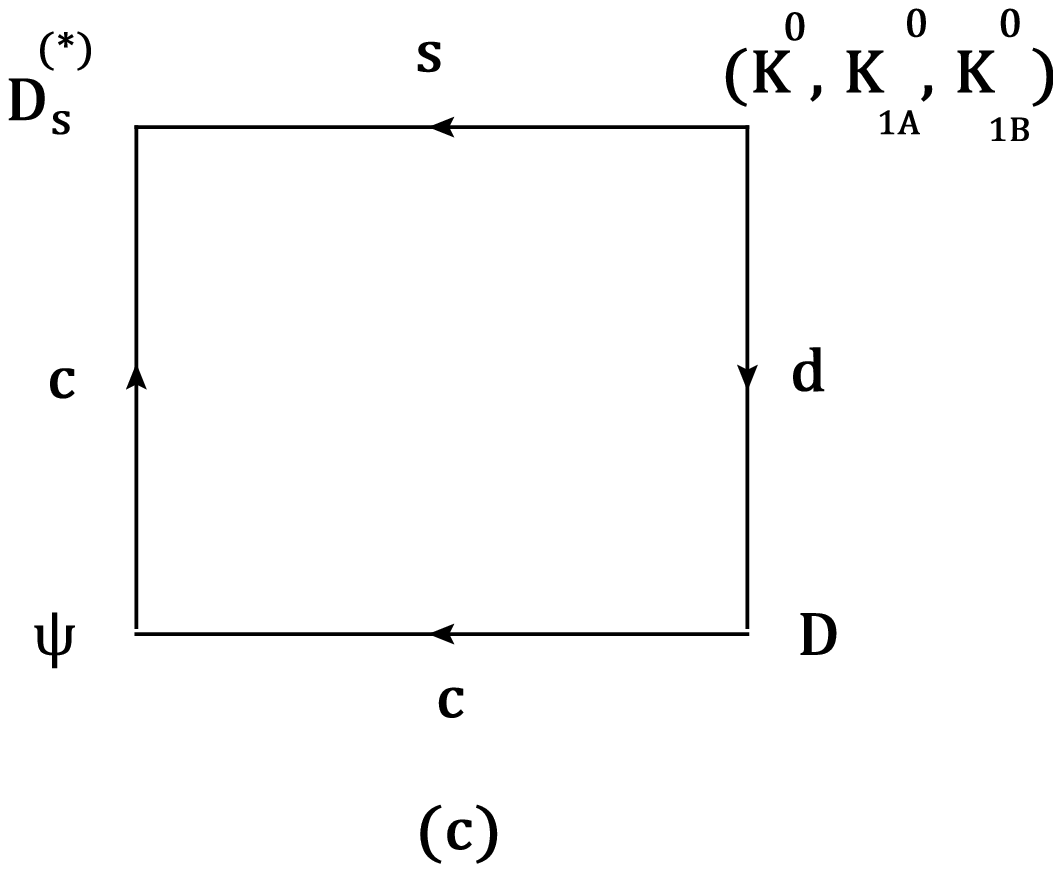}
\caption{4-particle diagrams show $(\psi, D^{(*)}, D, P)$ and $(\psi,  D^{(*)} D, A)$
vertices.} \label{4pv}
\end{figure}

Moreover, diagrams including  4 particles which  are considered in this paper are displaced in Fig (\ref{4pv}).
$(\psi, D^{_0}, D^{_{+}}, \pi^{_-})$, $(\psi, D^{_{*0}}, D^{_{+}}, \pi^{_-})$,
 $(\psi, D^{_0}, D^{_{+}}, a_{1}^{_-})$, $(\psi, D^{_0}, D^{_{+}},  b_{1}^{-})$, $(\psi, D^{_{*0}}, D^{_{+}}, a_{1}^{-})$
 and   $(\psi, D^{_{*0}}, D^{_{+}}, b_{1}^{_-})$  vertices can be explained via diagram (a).
Diagram (b) describes $(\psi, D^{_0}, \bar{D}^{_0}, \pi^{_0})$ and $(\psi, D^{_{*0}}, \bar{D}^{_{0}}, \pi^{_{0}})$
vertices while, $(\psi, D_{s}^{_{+}}, D^{_{-}}, K^{_{0}})$,
$(\psi, D_{s}^{_{*+}}, D^{_{-}}, K^{_0})$, $(\psi, D_{s}^{_{+}}, D^{_{-}}, K_{1A}^{_0})$,
 $(\psi, D_{s}^{_{+}},  D^{_{-}}, K_{1B}^{_0})$, $(\psi, D_{s}^{_{*+}}, D^{_{-}}, K_{1A}^{_0})$ and
  $(\psi, D_{s}^{_{*+}}, D^{_{-}}, K_{1B}^{_0})$   vertices are explained via diagram (c).

In the following  two subsections the strong couplings of  $(D^{(*)}, D^{(*)}, A)$, $(D^{(*)}, D^{(*)}, V)$, $(D_{1}, D_{1}, \pi)$
$(\psi,  D^{(*)}, D, P)$ and $(\psi,  D^{(*)}, D, A)$ vertices are derived.

\subsection{3-point functions and charm meson couplings}
In this section the $(D, D, A)$, $(D^{*}, D, A)$, $(D, D, V)$, $(D^{*}, D, V)$,
$(D^{*}, D^{*}, V)$ and $(D_{1}, D_{1}, P)$ vertices   couplings are  derived.
In our notation we use $D^{(*)}=D^{_{0(0*)}}, D^{_{\pm(\pm *)}}, D_{s}^{_{\pm(\pm*)}}$,
$A=a_{1}^{_-}, b_{1}^{_{-}}, K_{1A}^{_-}, K_{1B}^{_-},  K_{1A}^{_0}, K_{1B}^{_0}$,
 $V=(\rho^{_-}, K^{*})$ and  $P=(\pi^{_-}, \pi^{_0}, K^{0})$ for charm, axial vector,  vector and pseudoscalar mesons, respectively.
In this paper, the following definitions:
\begin{eqnarray}\label{gdda}
\langle D (p_{1}) |A(p_{2}, \varepsilon')\,D(p_{3})\rangle&=&2\,(\varepsilon'.p_{3})\,g_{_{DDA}},\nonumber\\
\nonumber\\
\langle D^{*} (p_{1}, \varepsilon) |A(p_{2}, \varepsilon')\,D(p_{3})\rangle&=&\big[
(\varepsilon^{*}.\varepsilon')\,(p_{3}.p_{1})-(\varepsilon^{*}.p_{3})\,(\varepsilon'.p_{1})\big.]\,g_{_{D^{*}\,D\,A}},\nonumber\\
\nonumber\\
\langle D (p_{1}) |V(p_{2}, \varepsilon)\,D(p_{3})\rangle&=&2\,(\varepsilon.p_{3})g_{_{DDV}},\nonumber\\
\nonumber\\
\langle D^{*} (p_{1}, \varepsilon_{1}) |V(p_{2}, \varepsilon_{2})\,D(p_{3})\rangle&=&\big[
(\varepsilon_{1}^{*}.\varepsilon_{2})\,(p_{3}.p_{1})-(\varepsilon_{1}^{*}.p_{3})\,(\varepsilon_{2}.p_{1})\big]\,g_{_{D^{*}\,D\,V}},\nonumber\\
\nonumber\\
\langle D^{*} (p_{1}, \varepsilon_{1}) |V(p_{2}, \varepsilon_{2})\,D^{*} (p_{3}, \varepsilon_{3})\rangle&=&
(\varepsilon^{*}_{1}.\varepsilon_{2})\,(\varepsilon_{3}.p_{3})\,g_{_{D^{*}\,D^{*}\,V}},\nonumber\\
\nonumber\\
\langle D_{1} (p_{1}, \varepsilon'_{1}) |P(p_{2})\,D_{1}(p_{3}, \varepsilon'_{2})\rangle&=&\big[
(\varepsilon_{1}^{'*}.\varepsilon_{2}^{'})\,(p_{3}.p_{1})-(\varepsilon_{1}^{'*}.p_{3})\,(\varepsilon'_{2}.p_{1})\big]\,g_{_{D_{1}\,D_{1}\,P}},
\end{eqnarray}
with $p_{1}=p_{2}+p_{3}$, are used for the  $(D, D, A)$, $(D^{*}, D, A)$, $(D, D, V)$, $(D^{*}, D, V)$,
$(D^{*}, D^{*}, V)$ and $(D_{1}, D_{1}, P)$
couplings \cite{Aliev1997, Belyaev1995, Bracco2012}.
Where as  emphasized in Eqs. (\ref{elementsdef3}) and (\ref{decayaxial}),   $\varepsilon$ denotes the polarization vector
of the vector meson $V$ and $D^{*}$  while  $\varepsilon'$ is used for axial vector mesons $A$ and $D_{1}$.

To obtain these  strong coupling
constants, we start with  the
correlation function including the currents of 3 considered particles. In  AdS/QCD approach
these 3-point functions
can be obtained by functionally differentiating
the $5$-D action with respect to their sources,
which are taken to be boundary values of the $5$-D fields
that have the correct quantum numbers as \cite{Witten1998, Gubser1998, Grigoryan2007, Abidin2008}
\begin{eqnarray}\label{3pddad}
\langle 0 |\mathcal{T}\left.\{ J_{A\parallel}^{\alpha a} (x) J_{A\perp}^{\mu b}(y)
	J_{A\parallel}^{\beta c}(w)\right.\}	| 0 \rangle&=&-\frac{\ \delta^{3} \rm{S}(DDA) \qquad }
	{\delta A_{\parallel\alpha}^{0a}(x) \,
	\delta A_{\perp\mu}^{0b}(y) \, \delta A_{\parallel\beta}^{0c}(w)}~~~~~~\rm({for~~ DDA ~~vertex}),  \\
\nonumber\\
\langle 0 |\mathcal{T}\left.\{ J_{V\perp}^{\mu a} (x) J_{A\perp}^{\nu b}(y)
	J_{A\parallel}^{\alpha c}(w)\right.\}	| 0 \rangle&=&-\frac{\ \delta^{3} \rm{S}(D^{*}DA) \qquad }
	{\delta V_{\perp\mu}^{0a}(x) \,
	\delta A_{\perp\nu}^{0b}(y) \, \delta A_{\parallel\alpha}^{0c}(w)} ~~~~~~\rm({for~~ D^{*}DA ~~vertex}),  \\
\nonumber\\
\langle 0 |\mathcal{T}\left.\{ J_{A\parallel}^{\alpha a} (x) J_{V\perp}^{\mu b}(y)
	J_{A\parallel}^{\beta c}(w)\right.\}	| 0 \rangle&=&-\frac{\ \delta^{3} \rm{S}(DDV) \qquad }
	{\delta A_{\parallel\alpha}^{0a}(x) \,
	\delta V_{\perp\mu}^{0b}(y) \, \delta A_{\parallel\beta}^{0c}(w)} ~~~~~~\rm({for~~ DDV ~~vertex}),  \\
\nonumber\\
\langle 0 |\mathcal{T}\left.\{ J_{V\perp}^{\mu a} (x) J_{V\perp}^{\nu b}(y)
	J_{A\parallel}^{\alpha c}(w)\right.\}	| 0 \rangle&=&-\frac{\ \delta^{3} \rm{S}(D^{*}DV) \qquad }
	{\delta V_{\perp\mu}^{0a}(x) \,
	\delta V_{\perp\nu}^{0b}(y) \, \delta A_{\parallel\alpha}^{0c}(w)} ~~~~~~\rm({for~~ D^{*}DV ~~vertex}),\\
\nonumber\\
\langle 0 |\mathcal{T}\left.\{ J_{V\perp}^{\mu a} (x) J_{V\perp}^{\nu b}(y)
	J_{V\perp}^{\sigma c}(w)\right.\}	| 0 \rangle&=&-\frac{\ \delta^{3} \rm{S}(D^{*}D^{*}V) \qquad }
	{\delta V_{\perp\mu}^{0a}(x) \,
	\delta V_{\perp\nu}^{0b}(y) \, \delta V_{\perp\sigma}^{0c}(w)} ~~~~~~\rm({for~~ D^{*}D^{*}V ~~vertex}),\\
\nonumber\\
\langle 0 |\mathcal{T}\left.\{ J_{A\perp}^{\mu a} (x) J_{A\perp}^{\nu b}(y)
	J_{A\parallel}^{\alpha c}(w)\right.\}	| 0 \rangle&=&-\frac{\ \delta^{3} \rm{S}(D_{1}D_{1}P) \qquad }
	{\delta A_{\perp\mu}^{0a}(x) \,
	\delta A_{\perp\nu}^{0b}(y) \, \delta A_{\parallel\alpha}^{0c}(w)} ~~~~~~\rm({for~~ D_{1}D_{1}P ~~vertex}),
\end{eqnarray}
where $\rm{S}(123)$ is  the relevant part of the $5$-D action for $(1, 2, 3)$ vertex.
To make a relation between the correlation functions  and their corresponding
vertexes,  we insert three complete sets of intermediate states with
the same quantum numbers as the meson currents into the correlation function. In the next step,
the matrix elements are defined in Eqs. (\ref{elementsdef3}), (\ref{decayaxial}) and
(\ref{decayp}) are used and the results can be obtained as:
 \begin{eqnarray}\label{ddaf}
&&\langle D (p_{1}) |A(p_{2}, \varepsilon')\,D(p_{3})\rangle=
\Omega_{\alpha\beta}(1, 3)\,\Lambda(D\,A\,D)\,
 \varepsilon'_{\mu}\,\hat{\mathcal{L}}\Bigg(\langle 0 |\mathcal{T}\left.\{ J_{A\parallel}^{\alpha a} (x) J_{A\perp}^{\mu b}(0)
	J_{A\parallel}^{\beta c}(w)\right.\}| 0 \rangle\Bigg),\nonumber\\
\nonumber\\
&&\langle D^{*} (p_{1}, \varepsilon) |A(p_{2}, \varepsilon')\,D(p_{3})\rangle=
\Omega_{\alpha}(3)\,\Lambda(D^{*}\,A\,D)\,
\,\varepsilon^{*}_{\mu}\,\varepsilon'_{\nu}\,\hat{\mathcal{L}}
\Bigg(\langle 0 |\mathcal{T}\left.\{ J_{V\perp}^{\mu a} (x) J_{A\perp}^{\nu b}(0)
	J_{A\parallel}^{\alpha c}(w)\right.\}	| 0 \rangle\Bigg),\nonumber\\
\nonumber\\
&&\langle D (p_{1}) |V(p_{2}, \varepsilon)\,D(p_{3})\rangle=
\Omega_{\alpha\beta}(1, 3)\,\Lambda(D\,V\,D)\,
 \varepsilon_{\mu}\,\hat{\mathcal{L}}\Bigg(\langle 0 |\mathcal{T}\left.\{ J_{A\parallel}^{\alpha a} (x) J_{V\perp}^{\mu b}(0)
	J_{A\parallel}^{\beta c}(w)\right.\}| 0 \rangle\Bigg),\,\nonumber\\
\nonumber\\
&&\langle D^{*} (p_{1}, \varepsilon_{1}) |V(p_{2}, \varepsilon_{2})\,D(p_{3})\rangle=
\Omega_{\alpha}(3)\,\Lambda(D^{*}\,V\,D)\,
\varepsilon^{*}_{1\mu}\,\varepsilon_{2\nu}\,\hat{\mathcal{L}}
\Bigg(\langle 0 |\mathcal{T}\left.\{ J_{V\perp}^{\mu a} (x) J_{V\perp}^{\nu b}(0)
	J_{A\parallel}^{\alpha c}(w)\right.\}	| 0 \rangle\Bigg),\nonumber\\
\nonumber\\
&&\langle D^{*} (p_{1}, \varepsilon_{1}) |V(p_{2}, \varepsilon_{2})\,D^{*} (p_{3}, \varepsilon_{3})\rangle=
\Lambda(D^{*}\,V\,D^{*})\,
\varepsilon^{*}_{1\mu}\,
\varepsilon_{2\nu}\,\varepsilon_{3\sigma}\hat{\mathcal{L}}
\Bigg(\langle 0 |\mathcal{T}\left.\{ J_{V\perp}^{\mu a} (x) J_{V\perp}^{\nu b}(0)
	J_{V\perp}^{\sigma c}(w)\right.\}	| 0 \rangle\Bigg),\nonumber\\
\nonumber\\
&&\langle D_{1} (p_{1}, \varepsilon'_{1}) |P(p_{2})\,D_{1}(p_{3}, \varepsilon'_{2})\rangle=
\Omega_{\alpha}(2)\,\Lambda(D_{1}\,P\,D_{1})\,
\,\varepsilon^{'*}_{1\mu}\,
\varepsilon'_{2\nu}\,\hat{\mathcal{L}}
\Bigg(\langle 0 |\mathcal{T}\left.\{ J_{A\perp}^{\mu a} (x) J_{A\perp}^{\nu b}(0)
	J_{A\parallel}^{\alpha c}(w)\right.\}	| 0 \rangle\Bigg),\nonumber
\end{eqnarray}
where,
\begin{eqnarray}
\hat{\mathcal{L}}=\int d^4x\, d^4w \, e^{ip_{1} x - ip_{3} w},\,~~~~~\Omega_{\alpha}(i)=\frac{p_{i\alpha}}{p_{i}^{2}},
\,~~~~~\Omega_{\alpha\,\beta}(i, j)=\Omega_{\alpha}(i)\,\Omega_{\beta}(j).
\end{eqnarray}
Moreover,
\begin{eqnarray}
\Lambda(\mathcal{O}_{1}\,\mathcal{O}_{2}\,\mathcal{O}_{3})=
\frac{(p_{1}^2 - m^{2}_{\mathcal{O}_{1}})}{f_{\mathcal{O}_{1}}}\,
\frac{(p_{2}^2 - m^{2}_{\mathcal{O}_{2}})}{f_{\mathcal{O}_{2}}}\,
\frac{(p_{3}^2 - m^{2}_{\mathcal{O}_{3}})}{f_{\mathcal{O}_{3}}},
\end{eqnarray}
is defined for the $\langle \mathcal{O}_{1}|\mathcal{O}_{2}\,\mathcal{O}_{3}\rangle$ matrix element. Moreover,
   in the final result,  the limit
 $(p_{1}^2, p_{2}^2, p_{3}^{2})\to (m^{2}_{\mathcal{O}_{1}}, m^{2}_{\mathcal{O}_{2}}, m^{2}_{\mathcal{O}_{3}})$ is
taken for considered  vertex.

Now  the relevant actions for every 3-point function are needed. For example,
to obtain $\rm{S}(DDA)$, we need to separate two pseudoscalar  fields (for $D$ mesons),
and one axial vector field (for $A$ meson) from three point action or for $\rm{S}(D^{*}DA)$,
we need a vector field, a pseudoscalar  field and one axial vector one. The results are calculated as
\begin{eqnarray}\label{3ddapf}
\rm{S}(DDA)&=&\int d^5x \bigg( \frac{l^{abc}}{z^{3}}  \left.[\frac{1}{2}A^{a}_{\mu}\,\partial^{\mu}(\pi^{b}\,\pi^{c})
-\partial_{\mu}(\pi^{a}-\phi^{a})A^{b}_{\mu}\,\pi^{c}\right.] \bigg),\\
S(D^{*}DA)&=&\int d^5x\bigg( \frac{f^{abc}}{2\,g_{5}^2\,z}\left.[\partial^{\mu}\phi^{b}\,V_{\mu\nu}\,A^{\nu c}\right.] +
\frac{h^{abc}}{z^3}\left.[V^{a\mu}\,A_{\mu}^{b}\pi^{c}\right.] +\frac{g^{abc}}{z^3}\left.[A_{\mu}^{a}\,V^{b\mu}\,
\pi^{c}\right.] \bigg)\\
\rm{S}(DDV)&=&\int d^5x \bigg( \frac{f^{abc}}{2g_5^2 z}\left.[\partial^\mu \phi^a V_{\mu\nu}^b\partial^\nu \phi^c
+ 2 \partial_z \partial_\nu \phi^a V_z^b\partial^\nu \phi^c \right.]+
 \frac{g^{abc}}{z^3} \left.[(\partial^\mu \pi^a - \partial^\mu \phi^a )
 V_\mu^b \pi^c- \partial_z \pi^a V_z^b \pi^c\right.]\bigg)\nonumber \\
&-&\int d^5x \bigg( \frac{h^{abc}}{2z^3}\left.[(\partial^\mu \left( \pi^a \pi^c \right)
- 2\partial^\mu \phi^a \pi^c ) V_\mu^b-\partial_z \left( \pi^a \pi^c \right)
V_z^b\right.]\bigg),\\
\rm{S}(D^{*}DV)&=&\int d^5x\bigg(\frac{k^{abc}}{z^3}\left.[V_{\mu}^{a}\,V^{\mu b}\,\pi^{c}+V_{z}^{a}\,V^{z b}\,\pi^{c}\right.]\bigg),\\
\rm{S}(D^{*}D^{*}V)&=&\int d^5x\bigg(\frac{f^{abc}}{g_{5}^2\,z}\left.[V_{\mu\nu}^{a}\,V^{\mu b}\,V^{\nu c}\right.]\bigg),\\
\rm{S}(D_{1}D_{1}P)&=&\int d^5x\bigg(\frac{l^{abc}}{z^3}\left.[A_{\mu}^{a}\,A^{\mu b}\,\pi^{ c}\right.]\bigg),
\end{eqnarray}
where
\begin{eqnarray}\label{gabc}
l^{abc}=-2\,i{\rm \,Tr\,} \Bigg.(\left\{t^a,X_0 \right\}
	\left\{ t^b, \left\{ t^c, X_0 \right\} \right\}\Bigg.),
\,~~~g^{abc}=-2i {\rm \,Tr\,} \Bigg.(\left\{t^a,X_0 \right\}
	\left[ t^b, \left\{ t^c, X_0 \right\} \right]\Bigg.),\\
h^{abc} =-2i {\rm \,Tr\,} \Bigg.(\left[t^a,X_0 \right]
	\left\{ t^b, \left\{ t^c, X_0 \right\} \right\}\Bigg.),\,~~~k^{abc} =-2i {\rm \,Tr\,} \Bigg.(\left[t^a,X_0 \right]
	\left[ t^b, \left\{ t^c, X_0 \right\} \right]\Bigg.).
\end{eqnarray}
In all of the actions  obtained here,  the $f^{abc}$ terms come from the gauge part
and the  terms containing  $l^{abc}$, $g^{abc}$, $h^{abc}$  and $k^{abc}$ are  from the chiral
part of the original action.  The  values of $f^{abc}$ are given in \cite{Mahmoud2013}
and for  $l^{abc}$, $h^{abc}$ and $k^{abc}$ the values which are used in numerical
 part of this paper,
are collected in Appendix.

It should be noted that  in $\rm{S}(DDA)$,  $\rm{S}(D^{*}DV)$ and $\rm{S}(D_{1}D_{1}P)$, the left hand gauge field term; ($L^{MN}\,L_{MN}$)
cancels the contribution of the right hand ones;  ($R^{MN}\,R_{MN}$); and in the final result,
the gauge part has no contribution.
Going  to Fourier transform space and using the relations \cite{Grigoryan20072,Abidin20083}:
\begin{align}
\phi^a(p,z) &= \phi^a(p^2,z) \frac{i p^\alpha}{p^2}
A_{\parallel\alpha}^{0a}(p)	\,,~~~~~~~~~~~~~~~~~\pi^a(p,z) = \pi^a(p^2,z) \frac{i p^\alpha}{p^2}
A_{\parallel\alpha}^{0a}(p)	\,,\label{furieh1} \\
A_{\perp \mu}^a(q,z) &= {\mathcal {A}}^a(q^2,z)\
A_{\perp \mu}^{0a}(q),\,~~~~~~~~~~~~~~~~~V_{\perp \mu}^b(q,z)= {\cal V}^b(q^2,z)\
V_{\perp \mu}^{0b}(q)	\,,\label{furieh2} \\
V_z^b(q,z) &= - \partial_z \tilde\pi^b(q^2,z)
\ \frac{i q^\alpha}{q^2}  V_{\parallel \alpha}^{0b}(q)	\,,~~~~~~~~~~~~~~~~~
\partial^\mu \to -i \left( {\rm relevant\ momentum} \right)^\mu\,\label{furieh3},
\end{align}
the strong couplings are obtained as:
\begin{eqnarray}
g_{_{DDA}}&=&g_{5}^{3}\,\int_{0}^{z_{0}} dz\, \bigg(\frac{\,l^{abc}}{z^3}\left.[\psi_{_{A}}^{a}( z)\,\pi^{b}(z)\,\pi^{c}(z)-
\,(\pi^{a}(z)-\phi^{a}(z))\psi_{_{A}}^{b}( z)\,\pi^{c}(z)\right.]\bigg),\label{gddafinal1}\\
g_{_{D^{*}DA}}&=&g_{5}\,\int_{0}^{z_{0}} dz\,\bigg( \frac{f^{abc}}{2\,z}\,
\left.[\psi_{_{V}}^{a}(z)\,
\phi^{b}(z)\,\psi_{_{A}}^{c}(z)\right.]+\frac{g_{5}^{2}\,g^{abc}}{\Delta_{1}\,z^3}\,
\left.[\psi_{_{A}}^{a}( z)\,
\psi_{_{V}}^{b}( z)\pi^{c}(z)\right]\bigg)\nonumber\\
&+&g_{5}^{3}\,\int_{0}^{z_{0}} dz\,\bigg(
\frac{h^{abc}}{\Delta_{1}\,z^3}\,\left[\psi_{_{V}}^{a}(z)\,
\psi_{_{A}}^{b}(z)\pi^{c}(z)\right]\bigg),\label{gddafinal2}\\
g_{_{DDV}}&=&g_{5}\,\int_{0}^{z_{0}} dz\,\bigg( \frac{f^{abc}}{2\,z}\,
\left.[
\phi^{a}(z)\,\psi_{_{V}}^{b}(z)\,
\phi^{c}(z)\right.]\,\Delta_{2}+\frac{g_{5}^{2}\,g^{abc}}{z^3}\,\left.[(\phi^a(z)-\pi^a(z))\,
\psi_{_{V}}^{b}(z)\,\phi^c(z)\right]\bigg)\nonumber\\
&+&g_{5}^{3}\,\int_{0}^{z_{0}} dz\,\bigg(
\frac{h^{abc}}{z^3}\,\left.[\phi^a(z)\,\psi_{_{V}}^{b}(z)\,\pi^c(z)\right.]\bigg),\label{gddafinal3}\\
g_{_{D^{*}DV}}&=&g_{5}^{3}\,\int_{0}^{z_{0}} dz\,\bigg(\frac{k^{abc}}{\Delta_{1}\,z^3}\left.[
\psi_{_{V}}^{a}(z)\,\psi_{_{V}}^{b}(z)\,\pi^c(z)\right]\bigg),\label{gddafinal4}\\
g_{_{D^{*}D^{*}V}}&=&g_{5}\,\int_{0}^{z_{0}} dz\,\bigg(\frac{f^{abc}}{\,z}\left.[
\psi_{_{V}}^{a}(z)\,\psi_{_{V}}^{b}(z)\,\psi_{_{V}}^c(z)\right]\bigg),\label{gddafinal5}\\
g_{_{D_{1}D_{1}P}}&=&g_{5}^{3}\,\int_{0}^{z_{0}} dz\,\bigg(\frac{l^{abc}}{\Delta_{1}\,z^3}\left.[
\psi_{_{A}}^{a}(z)\,\psi_{_{A}}^{b}(z)\,\pi^c(z)\right]\bigg),\label{gddafinal6}
\end{eqnarray}
where the parameters $\Delta_{1}$ and $\Delta_{2}$ are defined as
\begin{eqnarray}
\Delta_{1}=p_{1}^2+p_{3}^{2}-p_{2}^{2}, ~~~~~~~~~\Delta_{2}=p_{1}^2-p_{3}^2.
\end{eqnarray}
Note
that $\psi_{A}^{a}(z)$ and $\psi^{a}_{V}$ are dimensionless but  the units of $\phi^{a}(z)$ and $\pi^{a}(z)$ are $\rm{GeV}^{-1}$
 ( or in the units of $z$).
 So, $g_{_{D^{*}\,D\,A}}$,
and $g_{_{D^{*}\,D\,V}}$ and  $g_{_{D_{1}D_{1}P}}$ are in units  $\rm{GeV}^{-1}$ and other couplings are dimensionless.
\subsection{4-point functions and charm meson couplings}
In this subsection we consider $(\psi, D, D, P)$, $(\psi, D^{*}, D, P)$, $(\psi, D, D, A)$ and $(\psi, D^{*}, D, A)$
vertices.
To obtain these vertexes couplings, we start with the following 4-point functions:
\begin{eqnarray}
\langle 0 |\mathcal{T}\left.\{ J_{V\perp}^{\mu a} (x) J_{A\parallel}^{\alpha b}(y)\,J_{A\parallel}^{\beta c}(w)
	J_{A\parallel}^{\gamma d}(u)\right.\}	| 0 \rangle&=&\frac{i\, \delta^{4} \rm{S}(\psi DDP) \qquad }
	{\delta V_{\perp\mu}^{0a}(x) \,
	\delta A_{\parallel\alpha}^{0b}(y) \, \delta A_{\parallel\beta}^{0c}(w)\,\delta A_{\parallel\gamma}^{0d}(u) }
~~~~~~\rm({for~~ \psi DDP ~~vertex}),\label{4-1}  \\
\nonumber\\
\langle 0 |\mathcal{T}\left.\{ J_{V\perp}^{\mu a} (x) J_{V\perp}^{\nu b}(y)\,J_{A\parallel}^{\alpha c}(w)
	J_{A\parallel}^{\beta d}(u)\right.\}	| 0 \rangle&=&\frac{i\,\delta^{4} \rm{S}(\psi D^{*}DP) \qquad }
	{\delta V_{\perp\mu}^{0a}(x) \,
	\delta V_{\perp\nu}^{0b}(y) \, \delta A_{\parallel\alpha}^{0c}(w)\,\delta A_{\parallel\beta}^{0d}(u) }~~~~~~
\rm({for~~ \psi D^{*}DP ~~vertex}), \label{4-2}  \\
\nonumber\\
\langle 0 |\mathcal{T}\left.\{ J_{V\perp}^{\mu a} (x)  J_{A\parallel}^{\alpha b}(y)\,J_{A\parallel}^{\beta c}(w)
	J_{A\perp}^{\nu d}(u)\right.\}	| 0 \rangle&=&\frac{i\, \delta^{4} \rm{S}(\psi DDA) \qquad }
	{\delta V_{\perp\mu}^{0a}(x) \,
	\delta A_{\parallel\alpha}^{0b}(y) \, \delta A_{\parallel\beta}^{0c}(w)\,\delta A_{\perp\nu}^{0d}(u) }~~~~~~
\rm({for~~ \psi DDA ~~vertex}),\label{4-3}   \\
\nonumber\\
\langle 0 |\mathcal{T}\left.\{ J_{V\perp}^{\mu a} (x) J_{V\perp}^{\nu b}(y)\,J_{A\parallel}^{\alpha c}(w)
	J_{A\perp}^{\sigma d}(u)\right.\}	| 0 \rangle&=&\frac{i\, \delta^{4} \rm{S}(\psi D^{*}DA) \qquad }
	{\delta V_{\perp\mu}^{0a}(x) \,
	\delta V_{\perp\nu}^{0b}(y) \, \delta A_{\parallel\alpha}^{0c}(w)\,\delta A_{\perp\sigma}^{0d}(u) }~~~~~~\rm({for~~ \psi D^{*}DA ~~vertex}),\label{4-4}
\end{eqnarray}
where $(1, 2, 3, 4)$  vertex is described by the $\rm{S}(1234)$ part of the total action.
In this paper, the couplings
$g_{_{\psi\,D\,D\,P}}$, $g_{_{\psi\,D^{*}\,D\,P}}$, $g_{_{\psi\,D\,D\,A}}$ and $g_{_{\psi\,D^{*}\,D\,A}}$
couplings are defined as:
\begin{eqnarray}
\langle \psi (p_{1}, \varepsilon)| D (p_{2}) D(p_{3})\,P(p_{4})\rangle&=&(\varepsilon^{*}.q)\,g_{_{\psi\,D\,D\,P}},\label{psigdd}
\end{eqnarray}
\begin{eqnarray}
\langle \psi (p_{1}, \varepsilon)| D (p_{2}) D(p_{3})\,A(p_{4}, \varepsilon' )\rangle&=&
(\varepsilon^{*}.\varepsilon')\,g_{_{\psi\,D\,D\,A}},\label{psigdda}
\end{eqnarray}
\begin{eqnarray}
\langle \psi (p_{1}, \varepsilon_{1})| D^{*} (p_{2}, \varepsilon_{2}) D(p_{3})\,P(p_{4})\rangle&=&
(\varepsilon_{1}^{*}.\varepsilon_{2})\,g_{_{\psi\,D^{*}\,D\,P}},\label{psigdstardp}
\end{eqnarray}
\begin{eqnarray}
\langle \psi (p_{1}, \varepsilon_{1})| D^{*} (p_{2}, \varepsilon_{2}) D(p_{3})\,A(p_{4}, \varepsilon' )\rangle&=&
(\varepsilon_{1}^{*}.\varepsilon')\,(\varepsilon_{2}. p_{3})\,g_{_{\psi\,D^{*}\,D\,A}},\label{psigdstarda}
\end{eqnarray}
with $q=p_{3}+p_{4}=p_{1}-p_{2}$.

To obtain considered quartic vertices we insert four intermediate states in to the correlation functions  given in
Eqs. (\ref{4-1}-\ref{4-4}),  and then using  the definitions given in Eqs.
  (\ref{elementsdef3}), (\ref{decayaxial}) and
(\ref{decayp}), we obtain:
\begin{eqnarray}
&&\langle \psi (p_{1}, \varepsilon)| D (p_{2}) D(p_{3})\,P(p_{4})\rangle=
\Omega_{\alpha\beta\delta}(2, 3, 4)\,
\Lambda(\psi\,D\,D\,P)\,
\varepsilon^{*}_{\mu}\,\hat{\hat{\mathcal{L}}}
 \Bigg(\langle 0 |\mathcal{T}\left.\{ J_{V\perp}^{\mu a} (x) J_{A\parallel}^{\alpha b}(y)\,J_{A\parallel}^{\beta c}(0)
	J_{A\parallel}^{\delta d}(u)\right.\}	| 0 \rangle\Bigg),\nonumber\\
\nonumber\\
&&\langle \psi (p_{1}, \varepsilon)| D (p_{2}) D(p_{3})\,A(p_{4}, \varepsilon' )\rangle=
\Omega_{\alpha\beta}(2, 3)
\Lambda(\psi\,D\,D\,A)\,
  \varepsilon^{*}_{\mu}\,\varepsilon'_{\nu}\hat{\hat{\mathcal{L}}}\Bigg
 (\langle 0 |\mathcal{T}\left.\{ J_{V\perp}^{\mu a} (x)  J_{A\parallel}^{\alpha b}(y)\,J_{A\parallel}^{\beta c}(0)
	J_{A\perp}^{\nu d}(u)\right.\}	| 0 \rangle\Bigg),\nonumber\\
\nonumber\\
&&\langle \psi (p_{1}, \varepsilon_{1})| D^{*} (p_{2}, \varepsilon_{2}) D(p_{3})\,P(p_{4})\rangle
=\Omega_{\alpha\beta}( 3, 4)\,
\Lambda(\psi\,D^{*}\,D\,P)\,
\varepsilon^{*}_{1\mu}\,\varepsilon_{2\nu}\hat{\hat{\mathcal{L}}}
 \Bigg(\langle 0 |\mathcal{T}\left.\{ J_{V\perp}^{\mu a} (x) J_{V\perp}^{\nu b}(y)\,J_{A\parallel}^{\alpha c}(0)
	J_{A\parallel}^{\beta d}(u)\right.\}	| 0 \rangle\Bigg),\nonumber\\
\nonumber\\
&&\langle \psi (p_{1}, \varepsilon_{1})| D^{*} (p_{2}, \varepsilon_{2}) D(p_{3})\,A(p_{4}, \varepsilon' )\rangle=
\Omega_{\alpha}(3)\,\Lambda(\psi\,D^{*}\,D\,A)\,
  \varepsilon^{*}_{1\mu}\,\varepsilon_{2\nu}\,\varepsilon'_{\sigma}\hat{\hat{\mathcal{L}}}\Bigg
 (\langle 0 |\mathcal{T}\left.\{ J_{V\perp}^{\mu a} (x) J_{V\perp}^{\nu b}(y)\,J_{A\parallel}^{\alpha c}(0)
	J_{A\perp}^{\sigma d}(u)\right.\}	| 0 \rangle\Bigg),\nonumber
\end{eqnarray}
where
\begin{eqnarray}
\hat{\hat{\mathcal{L}}}=\int d^4x\, d^4y \,d^4u\, e^{ip_{1}x + ip_{2}y -ip_{4}u},\,~~~
\Omega_{\alpha\beta\delta}(i, j, k)= \Omega_{\alpha}(i)\,\Omega_{\beta}(j)\,\Omega_{\delta}(k).
\end{eqnarray}
For  $\langle \mathcal{O}_{1}|\mathcal{O}_{2}\,\mathcal{O}_{3}\,\mathcal{O}_{4}\rangle$ matrix element
the following definition is used:
\begin{eqnarray}
\Lambda(\mathcal{O}_{1}\,\mathcal{O}_{2}\,\mathcal{O}_{3}\,\mathcal{O}_{4})=
\frac{(p_{1}^2 - m^{2}_{\mathcal{O}_{1}})}{f_{\mathcal{O}_{1}}}\,
\frac{(p_{2}^2 - m^{2}_{\mathcal{O}_{2}})}{f_{\mathcal{O}_{2}}}
\,\frac{(p_{3}^2 - m^{2}_{\mathcal{O}_{3}})}{f_{\mathcal{O}_{3}}}\,
\frac{(p_{4}^2 - m^{2}_{\mathcal{O}_{4}})}{f_{\mathcal{O}_{4}}},
\end{eqnarray}
and   the limit
 $(p_{1}^2, p_{2}^2, p_{3}^{2}, p_{4}^{2})\to (m^{2}_{\mathcal{O}_{1}}, m^{2}_{\mathcal{O}_{2}}, m^{2}_{\mathcal{O}_{3}},
 m^{2}_{\mathcal{O}_{4}})$ is
applied in  the final result.

Now we need to obtain the relevant action for every correlation function. The results are obtained as:
\begin{eqnarray}
\rm{S}(\psi DDP)&=&\int d^5x\, \bigg( \frac{g^{abcd}}{2\,z^{3}}  \left.[\partial_{\mu}(\pi^{a}-\phi^{a})
\,V^{\mu b}\,\pi^{c}\,\pi^{d}+\partial_{z}\pi^{a}\,V^{b z}\,\pi^{c}\,\pi^{d}\right.]
+\frac{k^{abcd}}{z^{3}} \left.[\partial_{\mu}\phi^{a}\pi^{b}V^{c\mu}\pi^{d} \right]\bigg)\nonumber\\
&&-\int d^5x\,\bigg( \frac{l^{abcd}}{2\,z^{3}}
\left.[V^{a\mu}\,\pi^{b}\,\partial_{\mu}(\pi^{c}\pi^{d})+V^{a z}\,\pi^{b}\,\partial_{z}(\pi^{c}\pi^{d}) \right]
-\frac{h^{abcd}}{2\,z^{3}}  \left.[V^{a\mu}\partial_{\mu}\phi^{b}\pi^{c}
\pi^{d}\right.]\bigg)\nonumber\\
&&+\int d^5x\,\bigg(\frac{h^{abcd}}{6\,z^{3}} \left.[V^{a\mu}\partial_{\mu}(\pi^{b}\pi^{c}\pi^{d})
+V^{a z}\partial_{z}(\pi^{b}\pi^{c}\pi^{d})\right.]\bigg),\\
\rm{S}(\psi DDA)&=&\int d^5x\, \bigg(-\frac{g^{abcd}}{2\,z^{3}}  \left.[A^{a}_{\mu}\,V^{b\mu}\,\pi^{c}\pi^{d}\right.]
+\frac{h^{abcd}}{2\,z^{3}}  \left.[V^{a}_{\mu}\,A^{b\mu}\,\pi^{c}\pi^{d}\right.]+\frac{k^{abcd}}{z^{3}}
\left.[V^{a}_{\mu}\,\pi^{b}\,A^{c\mu}\,\pi^{d}\right.]\bigg),\\
\rm{S}(\psi D^{*}DP)&=&\int d^5x\, \bigg(\frac{f^{abcd}}{2\,g_{5}^2\,z}\left.[V^{a\mu} V^{b\nu} \partial_{\mu}\phi^{c}\partial_{\nu}\phi^{d}+V^{a\mu}\partial_{\nu}\phi^{b}V^{c}_{\mu} \partial^{\nu}\phi^{d}\right.]
-\frac{h^{abcd}}{2\,z^{3}}\left.[V^{a}_{ \mu}\,V^{b \mu}\pi^{c}\pi^{d}\right.]\bigg)\nonumber\\
&&-\int d^5x\, \bigg(\frac{h^{abcd}}{2\,z^{3}}\left.[V^{a}_{z}\,V^{b z}\pi^{c}\pi^{d}\right.]\bigg),\\
\rm{S}(\psi D^{*}DA)&=&\int d^5x\, \bigg(\frac{f^{abcd}}{2\,g_{5}^2\,z}\left.[V^{a}_{\mu}\,A^{b}_{\nu}V^{c\mu}\,\partial^{\nu}\phi^{d}
+V^{a}_{\mu}\,\,A^{b}_{\nu}\,\partial^{\mu}\phi^{c}\,V^{d\nu}\,\right]\bigg),
\end{eqnarray}
with
\begin{eqnarray}
f^{abcd}&=&2{\rm \,Tr\,} \Bigg.(\left[t^{a}, t^{b}\right]\left[t^{c}, t^{d}\right]\Bigg.),\nonumber\\
g^{abcd}&=&2{\rm \,Tr\,} \Bigg.(\left\{t^{a},X_0 \right\}\left[t^{b}, \left\{t^{c}, \left\{t^{d},X_0 \right\} \right\}\right]\Bigg.),\nonumber\\
h^{abcd}&=&2{\rm \,Tr\,}\Bigg.(\left[t^{a},X_0 \right]\left\{t^{b}, \left\{t^{c}, \left\{t^{d},X_0 \right\} \right\}\right\}\Bigg.),\nonumber\\
k^{abcd}&=&2{\rm \,Tr\,} \Bigg.(\left\{t^{a},\left\{t^{b}, X_0\right\} \right\}\left[t^{c},\left\{t^{d}, X_0\right\} \right]\Bigg.),\nonumber\\
l^{abcd}&=&2{\rm \,Tr\,} \Bigg.(\left[t^{a},\left\{t^{b}, X_0\right\} \right]\left\{t^{c},\left\{t^{d}, X_0\right\} \right\}\Bigg.),
\end{eqnarray}
where $f^{abcd}$ can be written in terms of structure constant as
$f^{abcd}=-f^{abm}\,f^{cdm}$. The  values of
$g^{abcd}$, $h^{abcd}$, $k^{abcd}$ and $l^{abcd}$  used in this paper
 are presented in Appendix. Using Eqs. (\ref{furieh1}, \ref{furieh2}) and (\ref{furieh3}),
  and  then by functional derivation according to
Eqs. (\ref{4-1}- \ref{4-4}), the final results for $g_{_{\psi\,D\,D\,P}}$,
 $g_{_{\psi\,D^{*}\,D\,P}}$, $g_{_{\psi\,D\,D\,A}}$
and $g_{_{\psi\,D^{*}\,D\,A}}$  couplings are obtained as:

\begin{eqnarray}\label{gddafinal}
&&g_{_{\psi\,D\,D\,P}}=g_{5}^{4}\,\int_{0}^{z_{0}} dz\,
\bigg(\frac{g^{abcd}}{2\,z^{3}} \left.[(\pi^{a}(z)-\phi^{a}(z))\psi_{_{V}}^{b}(z)\,\pi^{c}(z)\,\pi^{d}(z) \right.]
-\frac{l^{abcd}}{2\,z^{3}}\left.[\psi_{_{V}}^{a}(z)\,\pi^{b}(z)\,\pi^{c}(z)\,\pi^{d}(z) \right.]\bigg)\nonumber\\
&&~~~~~~~~~~~+g_{5}^{4}\,\int_{0}^{z_{0}} dz\,\bigg( \frac{h^{abcd}}{6\,z^{3}}
\left.[\psi_{_{V}}^{a}(z)\,(3\,\phi^{b}(z)+\pi^{b}(z)) \,\pi^{c}(z)\,\pi^{d}(z)\right.]+
\frac{k^{abcd}}{z^{3}}\left.[ \phi^{a}(z)\, \pi^{b}(z)\,
\psi_{_{V}}^{c}(z)\,\pi^{d}(z)\right.]\bigg),\\
&&g_{_{\psi\,D\,D\,A}}=g_{5}^{4}\,
\int_{0}^{z_{0}} dz\,\bigg(-\frac{g^{abcd}}{2\,z^{3}}\left.[\psi_{_{A}}^{a}(z)\,
\psi_{_{V}}^{b}(z)\,\pi^{c}(z)\,\pi^{d}(z)\right.]
+\frac{h^{abcd}}{2\,z^{3}}
\left.[\psi_{_{V}}^{a}( z)\,\psi_{_{A}}^{b}(z)\,\phi^{c}(z)\,\phi^{d}(z)\right.]\bigg)\nonumber\\
&&~~~~~~~~~~~-g_{5}^{4}\,\int_{0}^{z_{0}} dz\,\bigg(\frac{k^{abcd}}{z^{3}}
\left.[\psi_{_{V}}^{a}(z)\,\phi^{b}(z)\,\psi_{_{A}}^{c}(z)\,\pi^{d}(z)\right.]\bigg),\\
&&g_{_{\psi\,D^{*}\,D\,P}}=g_{5}^{2}\,
\int_{0}^{z_{0}} dz\,\bigg(\frac{f^{abcd}}{4\,z}\,
\left.[\psi_{_{V}}^{a}(z)\,\phi^{b}(z)\,\psi_{_{V}}^{c}(z)\,\phi^{d}(z)\right.]\,\Delta_{3}
-\frac{g_{5}^{2}\,h^{abcd}}{2\,z^{3}}\left.[\psi_{_{V}}^{a}( z)\,
\psi_{_{V}}^{b}(z)\,\pi^{c}(z)\,\pi^{d}(z)\right.]\bigg),\\
&&g_{_{\psi\,D^{*}\,D\,A}}=g_{5}^{2}\,
\int_{0}^{z_{0}} dz\,\bigg(\frac{f^{abcd}}{2\,z}\left.[\psi_{_{V}}^{a}(z)\,
\psi_{_{A}}^{b}(z)\,\psi_{_{V}}^{c}(z)\,\phi^{d}(z)\right.]\bigg),
\end{eqnarray}
where
\begin{eqnarray}
\Delta_{3}=q^2-p_{3}^{2}-p_{4}^{2}.
\end{eqnarray}
Here,  the strong couplings of $\psi\,D\,D\,P$ and $\psi\,D^{*}\,D\,A$ vertices are in the units
of $\rm{GeV}^{-1}$ while  $g_{_{\psi\,D\,D\,A}}$ and $g_{_{\psi\,D^{*}\,D\,P}}$ are dimensionless.

\section{Numerical analysis}\label{sec.num}
In this section, our numerical analysis is presented for the strong
coupling constants $g_{_{D D A}}$ and $g_{_{D^{*} D A}}$, $g_{_{D\,D\,V}}$,
 $g_{_{D^{*}\,D\,V}}$,  $g_{_{D^{*}\,D^{*}\,V}}$,  $g_{_{D_{1}\,D_{1}\,P}}$,
 $g_{_{\psi\,D\,D\,P}}$,
 $g_{_{\psi\,D^{*}\,D\,P}}$, $g_{_{\psi\,D\,D\,A}}$
and $g_{_{\psi\,D^{*}\,D\,A}}$.
In the first step of numerical analysis we must determine the values of $z_{0}$, $m_{q}$ and $\sigma_{q}$
for $q=(u, d, s, c)$ using experimental values of the masses.

The values of the experimental masses are utilized to fit $z_{0}$, quark masses and
quark condensates are presented in Table \ref{expmass}.

\begin{table}[th]
\caption{The experimental values of mass are used to fit $z_{0}$, $m_{u, d, s, c}$
and $\sigma_{u, d, s, c}$. These values are taken from \cite{pdg}. } \label{expmass}
\begin{tabular}{|c||c|c||c|c||c|}
\hline
Meson  &Mass ~(MeV)&Meson  &Mass ~(MeV)&Meson  &Mass ~(MeV)\\
\hline
&&&&&\\
$\rho^{0}$&$775.49\pm 0.34$&$\pi^{0}$&$134.97\pm 0.00$&$K^{_{*-}}$&$891.66\pm 0.26$\\
&&&&&\\
$\rho^{-}$&$775.40\pm 0.34$&$\pi^{-}$&$139.75\pm0.00$&$D^{-}$&$1869.65\pm 0.05$\\
&&&&&\\
$a_{1}^{-}$&$1230\pm 40$&$K^{-}$&$493.67\pm 0.01$&$D^{_{*-}}$&$2010.26\pm 0.05$\\
&&&&&\\
\hline
\end{tabular}
\end{table}
To evaluate $z_{0}$,
the observable which
does not depend on any other parameter  is used.  For this purpose, we can use the vector mesons with
${M_V^a}^2 =0$. Our choice in this part is the mass of the $\rho^{0}$ meson which gives us
$z_{0}^{-1}=(323 \pm 1)\,\rm{MeV}$.

After estimating $z_{0}$, we use the masses of the light mesons $\rho^{-}$, $a_{1}^{-}$,
$\pi^{0}$ and $\pi^{-}$ to fit $(m_{u}, m_{d}, \sigma_{u}, \sigma_{d}$).
In addition, $(m_{s}, \sigma_{s})$ are determined using the experimental masses of
the strange states
$K^{-}$ and $K^{_{*-}}$. Finally,  the experimental values of $m_{D^{-}}$ and $m_{D^{_{*-}}}$
are utilized to find fitted values of $(m_{c}, \sigma_{c})$.
Numerically,
 the best global fit  for the parameters $m_{q}$ in $\rm{MeV}$ are
 obtained as: $m_{u}=(8.5 \pm  2.5)$, $m_{d}=(12.36 \pm  2.45)$, $m_{s}=(195.31 \pm 5.89)$ and
 $m_{c}=(1590.56 \pm 8.42)$. Moreover, for  the quark condensates $\sigma_{q}$ in $\rm{MeV}^{3}$
 the best global fit values are $\sigma_{u}=(173.65 \pm 2.21)^{3}$, $\sigma_{d}=(177.42 \pm 3.15)^{3}$,
  $\sigma_{s}=(226.20 \pm 3.72)^{3}$ and
 $\sigma_{c}=(310.35 \pm 5.65)^{3}$.

Having all of these parameters in hand, we can estimate the pseudoscalar, vector and axial
vector meson masses. Table \ref{mesonmass} includes our predictions and   the experimental
values of the mesons which  are given  taken from \cite{pdg, pdg2}.
As it can be seen from the masses  reported in Table \ref{mesonmass},
the uncertainty  for $\psi$ and $\omega$ meson masses are lower than the those for the others.
For these two vector mesons, the uncertainties comes from $z_{0}$ parameter, while for the other
mesons, the quark masses and quark condensates are also included in the lower and higher
bounds of the masses. The mass of the $K_{1A}^{-}$ state is estimated using sum rules in \cite{Kwei} as
 $m_{K_{1A}^{-}}=(1310\pm 60)~\rm{MeV}$ while,  our analysis  predicts    $m_{K_{1A}^{-}}=(1316.52\pm 7.50)~\rm{MeV}$.
\begin{table}[th]
\caption{Global fit  to meson's masses as well as the experimental
values are reported in \cite{pdg, pdg2}.  } \label{mesonmass}
\begin{tabular}{|c||c|c|c||c|c|}
\hline
Meson  &Mass ~(MeV)&This work~(MeV)&Meson  &Mass ~(MeV)&This work ~(MeV)\\
\hline
&&&&&\\
$D^{_{*0}}$&$2006.85\pm 0.05$&$2005.53\pm6.65$&$D^{_{0}}$&$1864.83\pm 0.05$&$1861.50\pm 3.58$\\
&&&&&\\
$D_{s}^{_{*-}}$&$2112.20\pm 0.40$&$2122.90\pm 9.42$&$K^{0}$&$497.61\pm 0.01$&$499.21\pm 1.82$\\
&&&&&\\
$\omega$&$782.65\pm 0.12$&$779.45\pm 0.12$&$D_{s}^{_{-}}$&$1968.34\pm 0.07$&$1972.63\pm 2.37$\\
&&&&&\\
$\psi$&$3096.90\pm 0.00$&$3095.20\pm 0.15$&$\eta_{c}$&$2983.90 \pm 0.50$&$2979.62 \pm 2.43$\\
&&&&&\\
$D_{1}^{_{0}}$&$2420.80\pm 0.50$&$2423.62\pm 4.52$&$D_{s1}^{_{-}}$&$2459.50 \pm 0.60$&$2461.50\pm 5.42$\\
&&&&&\\
$D_{1}^{_{-}}$&$2423.40\pm 3.10$&$2427.25\pm 3.28$&$\chi_{_{c1}}$&$3510.67\pm 0.05$&$3507.28\pm 5.25$\\
&&&&&\\
\hline
\end{tabular}
\end{table}

Our prediction for the decay constants  of some  mesons are presented in Table \ref{mesondecay}.
The experimental measurements of the considered decay constants are also given in this table.
The measured
values for $f_{D^{-}}$ and  $f_{D_{s}^{-}}$ are averages from
lattice QCD results,  taken from Ref. \cite{pdg}. The decay constants of  $\rho^{-}$  and $a_{1}^{-}$
 mesons are taken from   \cite{Donoghue2014} and  \cite{Isgur1989}, respectively.
The other measured values are taken
from experimental data.

\begin{table}[th]
\caption{Our predictions
 for the  decay constants of nine selected mesons. The measured
value are taken from \cite{pdg, Donoghue2014, Isgur1989}} \label{mesondecay}
\begin{tabular}{|c||c|c|c||c|c|c||c|}
\hline
Observable  &Measured (MeV)&This work (MeV)&Observable&Measured (MeV)&This work (MeV)
&Observable&This work (MeV)\\
\hline
&&&&&&&\\
$f_{\pi^{-}}$&$92.07\pm 1.20$&$97.16\pm 2.63$&$f_{a_{1}^{-}}^{1/2}$&$420\pm 40$&$415.21\pm 4.62$
&$f_{D^{_{*-}}}$&$573.05\pm 3.42$\\
&&&&&&&\\
$f_{K^{-}}$&$110\pm 0.30$&$103.17\pm 4.91$&$f_{D^{-}}$&$149.80\pm 0.80$&$158.35\pm 4.16$
&$f_{D_{s}^{_{*-}}}$&$534.92\pm 5.82$\\
&&&&&&&\\
$f_{\rho^{-}}^{1/2}$&$345\pm 8$&$336.89\pm 2.13$&$f_{D_{s}^{_{-}}}$&$176.10\pm 0.80$&$167.66\pm 4.85$
&$f_{D_{1}^{_{-}}}$&$712.06\pm 9.81$\\
&&&&&&&\\
\hline
\end{tabular}
\end{table}

It should be noted that in our model,  there are no differences  between the mass and decay constants
of  $a_{1}^{-}$ and $b_{1}^{-}$. In addition, the mass and the decay constants of $K_{1A}^{-}$
and $K_{1A}^{0}$  are similar to the values   obtained for  $K_{1B}^{-}$
and $K_{1B}^{0}$, respectively.

\begin{figure}[!th]
 \label{wavefplot}
\includegraphics[width=8cm,height=7cm]{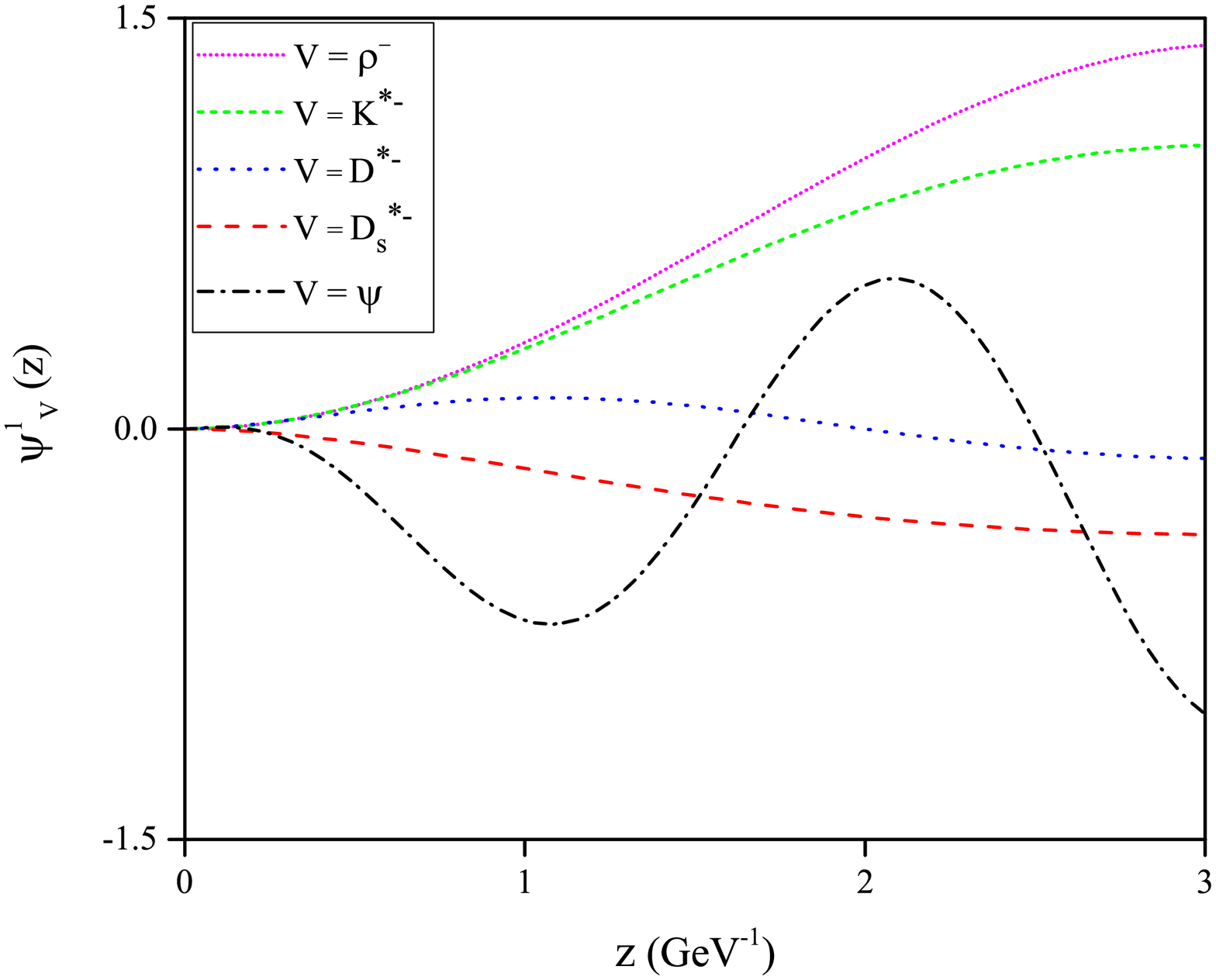}
\includegraphics[width=8cm,height=7cm]{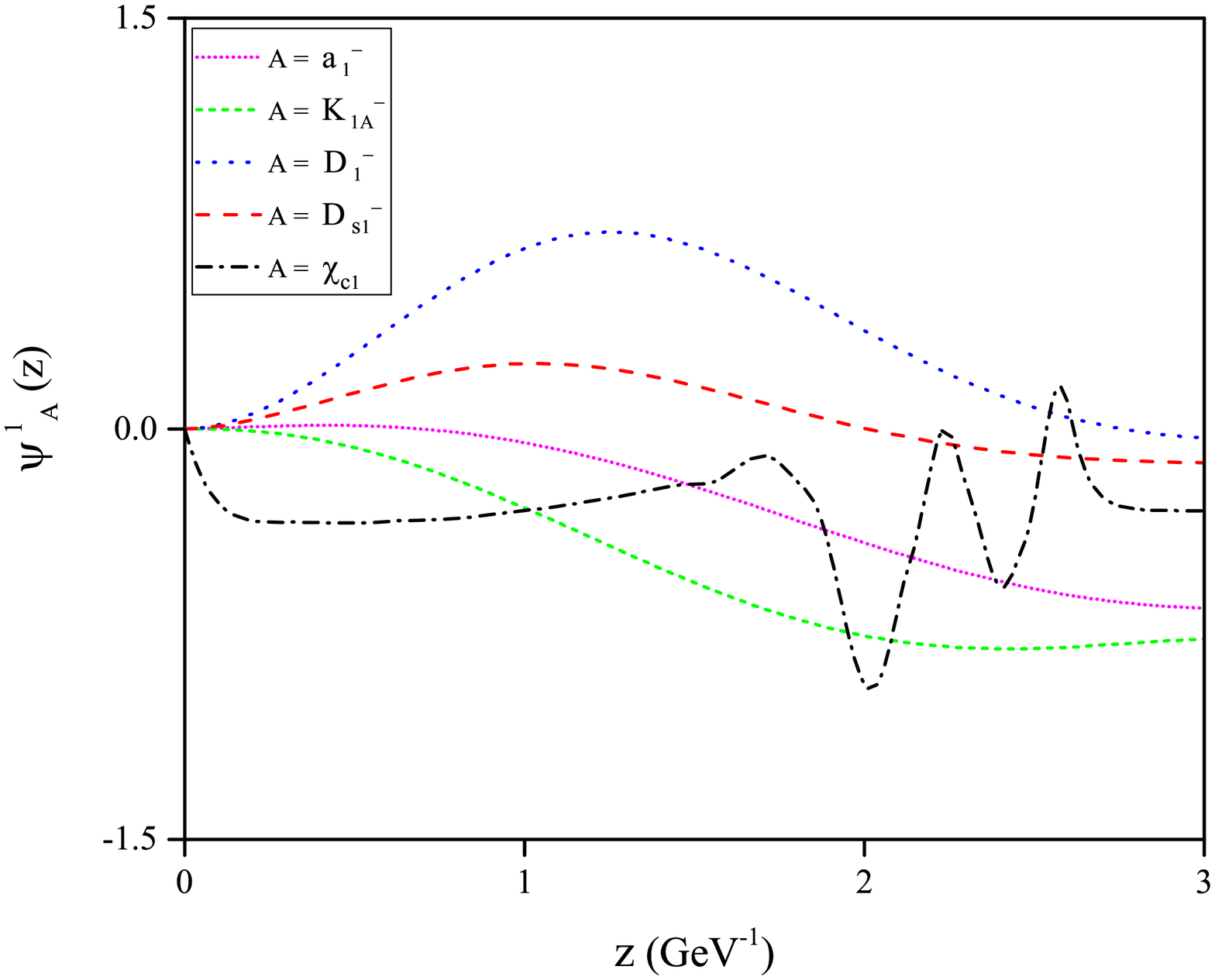}
\includegraphics[width=8cm,height=7cm]{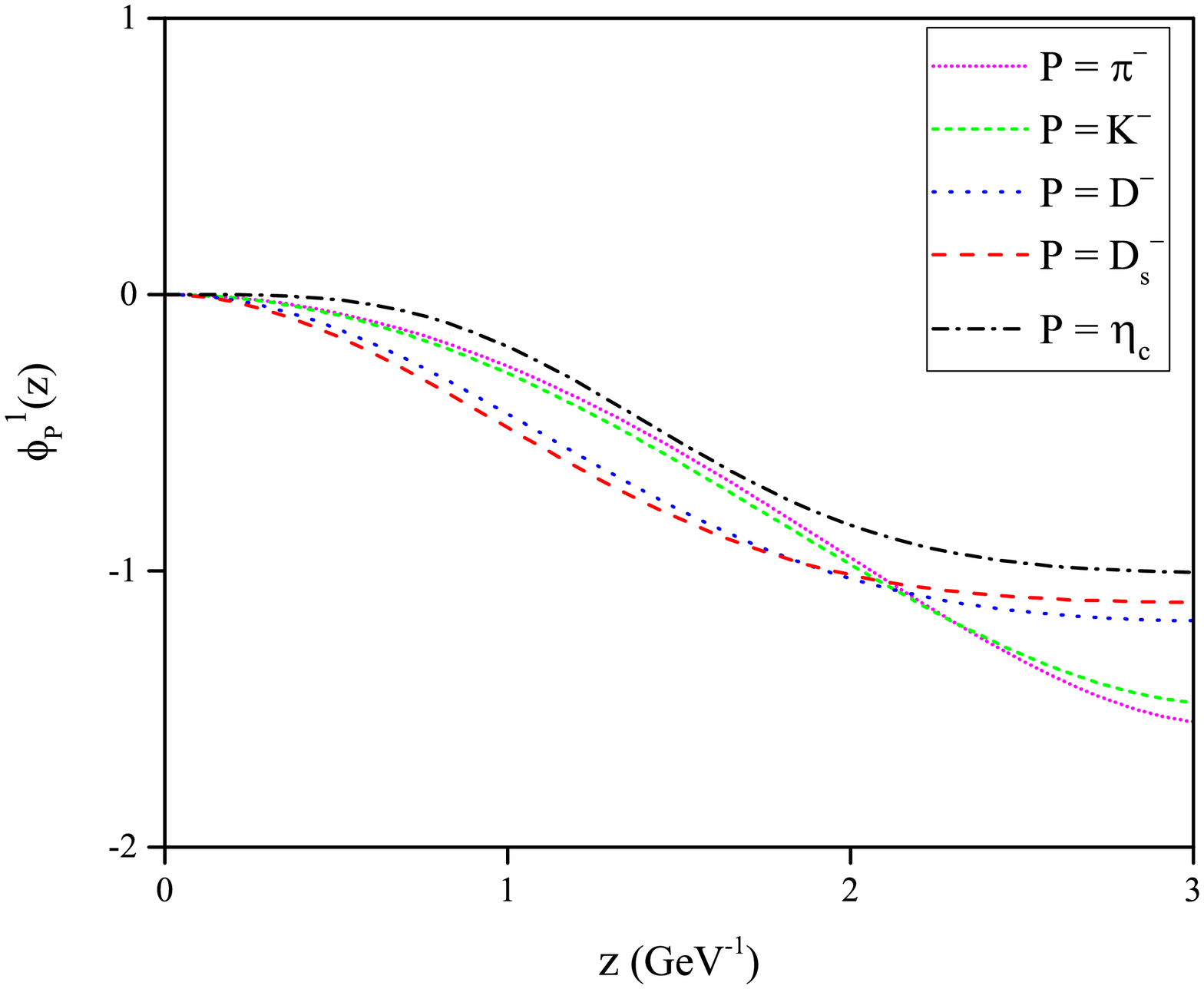}
\includegraphics[width=8cm,height=7cm]{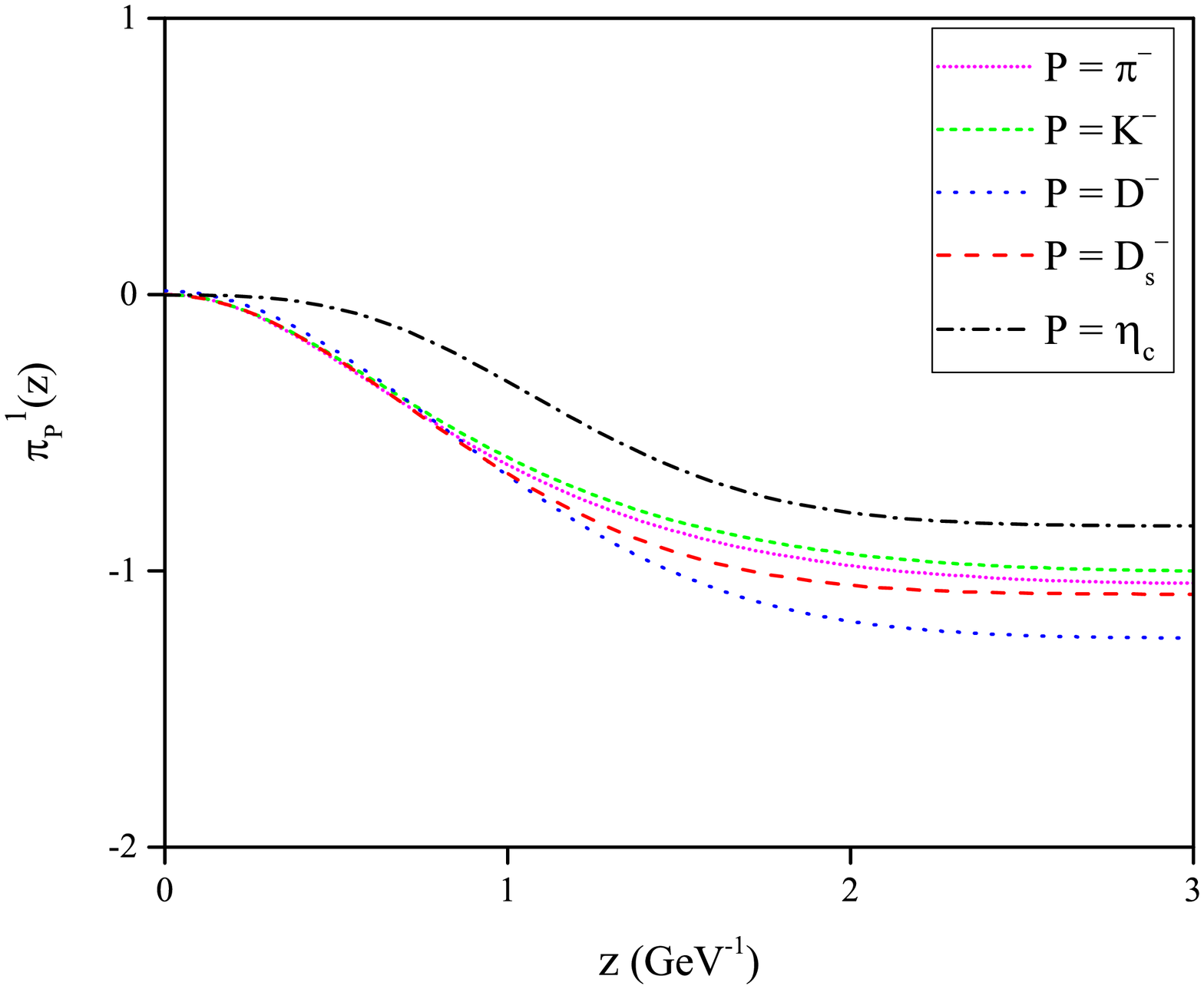}
\caption{Plots of the wave functions $\psi^{1}_{V}(z)$, $\psi^{1}_{A}(z)$, $\phi^{1}_{P}(z)$ and $\pi^{1}_{P}(z)$
 for $V=(\rho^{-}, K^{_{*-}}, D^{_{*-}}, D_{s}^{_{*-}}, \psi)$,
 $A=(a_{1}^{_{-}}, K_{1A}^{_{-}}, D_{1}^{_{-}}, D_{s1}^{_{-}}, \chi_{_{c1}})$ and
 $P=(\pi^{_{-}}, K^{_{-}}, D^{_{-}}, D_{s}^{_{-}}, \eta_{c})$ as functions of the radial coordinate $z$ in the interval $(\varepsilon, z_{0})$ . }
\end{figure}

Now  the wave functions for the studied mesons can be evaluated.
The wave functions $\psi^{1}_{V}$, $\psi^{1}_{A}$, $\phi^{1}_{P}$ and $\pi^{1}_{P}$
 as  functions of $z$ are plotted in  Fig. \ref{wavefplot} for $\varepsilon\leq z\leq z_{0}$.
 Here,  $\rho^{-}, a_{1}^{-}$ and  $\pi^{-}$ are selected from  the light mesons while, from the strange mesons
we plot the wave functions for  $K^{*-}, K_{1A}^{-}$ and $K^{-}$. Moreover, from the charm mesons group the plots are drawn
for  $D^{*-}, D_{1}^{-}$ and  $D^{-}$  states and  the mesons  $(D_{s}^{*-}, D_{s1}^{-}, D_{s}^{-})$
and  ($\psi, \chi_{_{c1}}, \eta_{c}$) are chosen from the charmed-strange and    $q\bar{q}$ states, respectively.

In this figure for the light,  strange,
charm, charm-strange
and  $q\bar{q}$ mesons, the plots are
displaced
with short-dot, short- dash, dot, dash and dash- dotted lines, respectively.
For $(\pi^{-}, \rho^{-}, a_{1}^{-}, K^{-}, K^{*-}, D^{-}, D^{*-})$ the valuse of the masses, taken
 from the experimental data  are reported in Table  \ref{expmass}
 while, for the other  ground state  mesons,  the  masses  are taken form our predictions
 given
in Table \ref{mesonmass}.

It should be noted that, since the values of $(m_{u}, \sigma_{u})$ are close to those of  $(m_{d}, \sigma_{d})$,
and the masses of $D^{*0}$ and $D^{*-}$ have almost no differences, the plot of $\psi_{D^{*0}}$
is similar to the    $\psi_{D^{*-}}$. Similarities of   plots of $\rho^{0}$, $\pi^{0}$, $D_{1}^{0}$, $D^{0}$ and  $K_{1A}^{0}$
are similar to those  obtained for $\rho^{-}$, $\pi^{-}$, $D_{1}^{-}$, $D^{-}$ and  $K_{1A}^{-}$, respectively.
For this reason,  in Fig. \ref{wavefplot} just one of these two choices are displaced.

Now we move to 3-particle states couplings defined in Eqs.
(\ref{gddafinal1}-  \ref{gddafinal6}). In this paper, to evaluate charm meson couplings to the axial vector mesons,
the mass of $b_{1}^{-}$ is taken from PDG as $m_{b_{1}^{-}}=(1229.50 \pm 3.20)~ \rm{MeV}$ \cite{pdg}.
Moreover, for $K_{1B}^{-}$, the mass is taken from 3PSR prediction as  $m_{K_{1B}^{-}}=(1340 \pm 80)~ \rm{MeV}$
\cite{Kwei}.
Our predictions for $g_{_{D\,D\,A}}$, $g_{_{D^{*}\,D\,A}}$
$g_{_{D\,D\,V}}$,   $g_{_{D^{*}\,D\,V}}$, $g_{_{D^{*}\,D^{*}\,V}}$ and $g_{_{D_{1}\, D_{1}\, P}}$ are reported in Tables \ref{couplingsf1} and \ref{couplingsf2}.
Notice that the main uncertainty  in the values of the  couplings comes from
$\sigma_{q}, ~(q=u, d, s, c)$ and the meson masses.

\begin{figure}
\includegraphics[width=7.5cm,height=6.5cm]{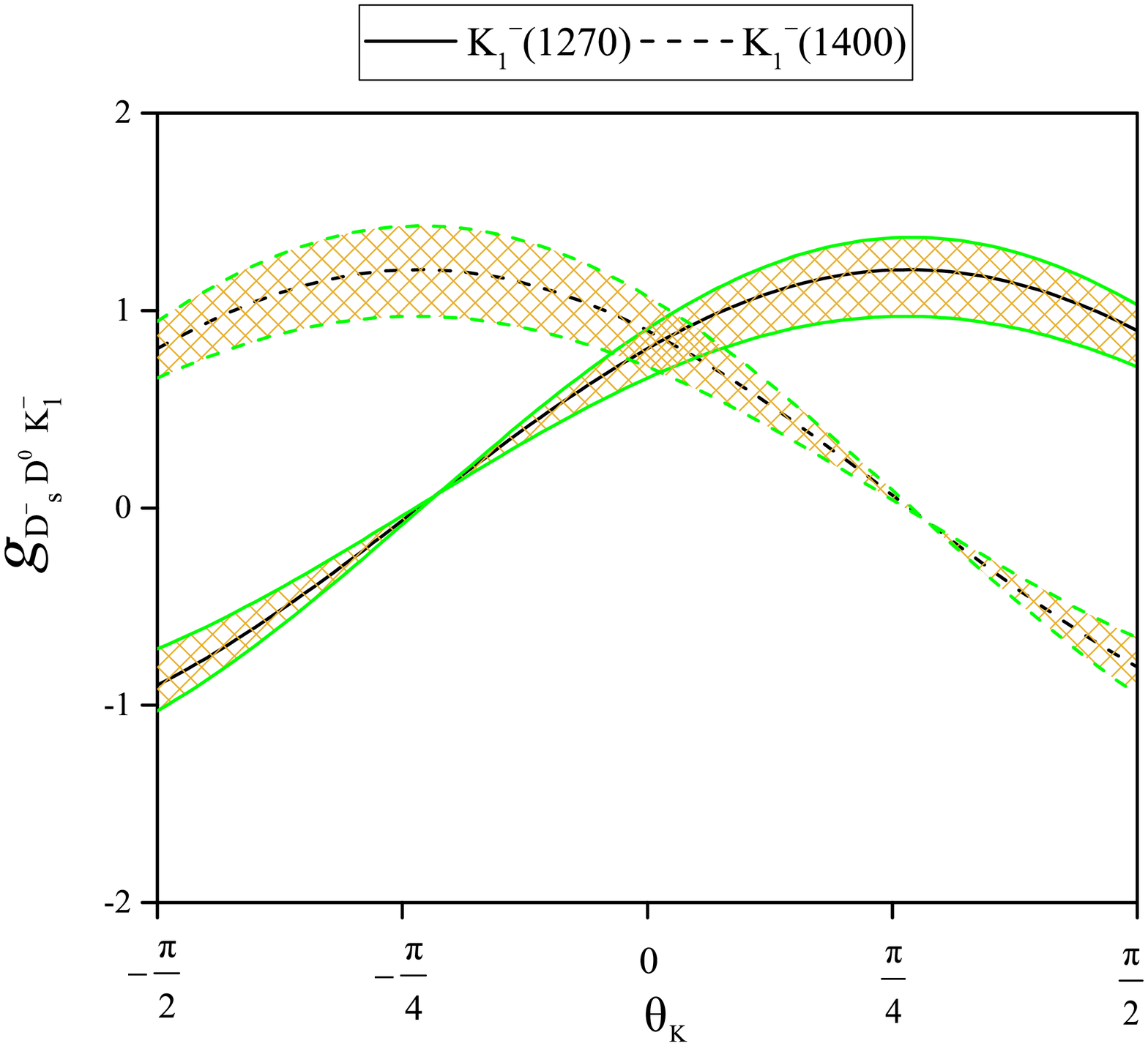}
\includegraphics[width=7.5cm,height=6.5cm]{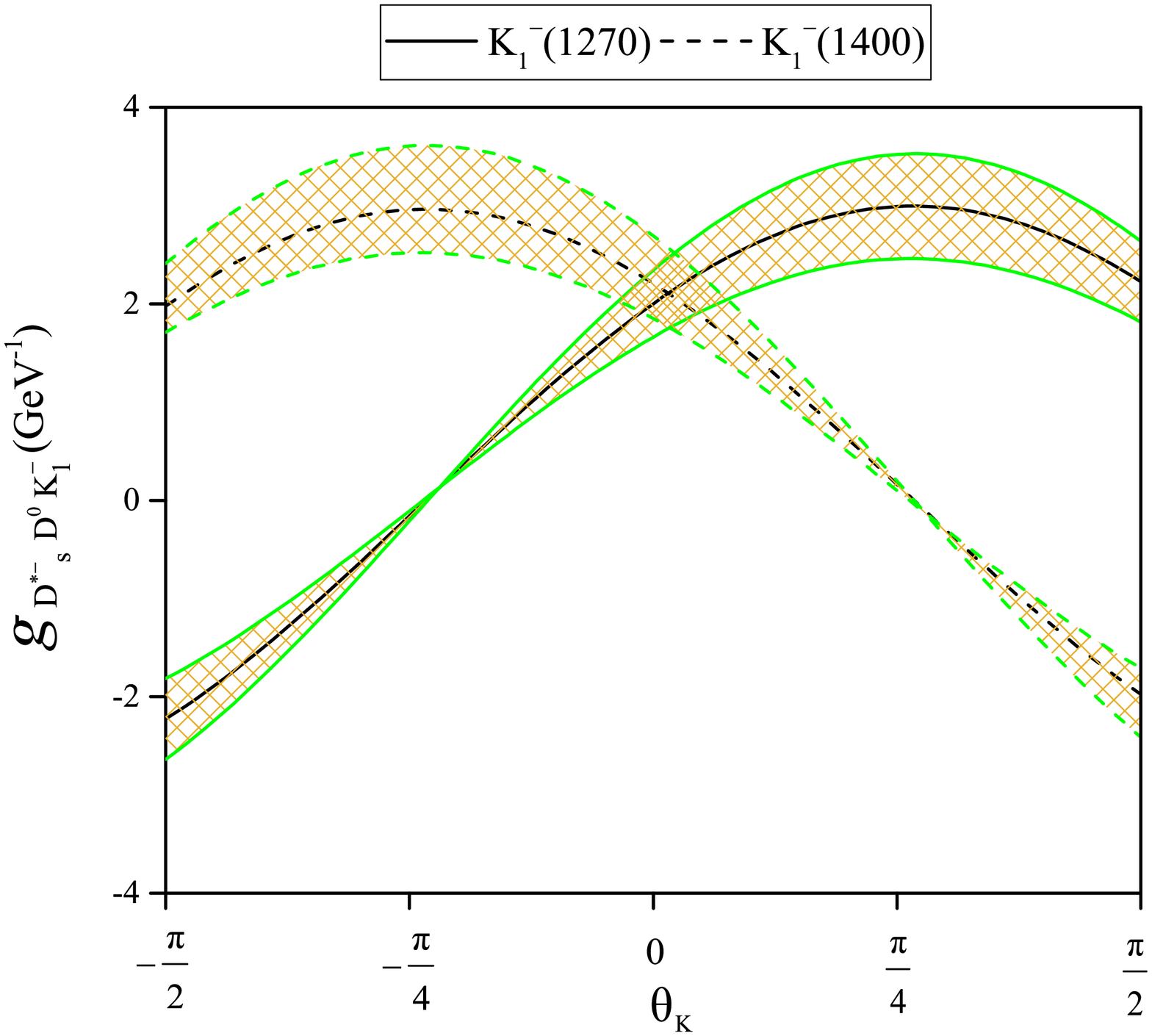}
\caption{The strong coupling constants $g_{D_s^{-} D^{0} K_1^{-}}$ and $g_{D_s^{*-} D^{0} K_1^{-}}$
for $K_1= K_{1}(1270), K_1(1400)$ as a function of the mixing
angle $\theta_{K}$ as well as the uncertainty regions.   } \label{fandgk1}
\end{figure}

\begin{table}[th]
\caption{Our predictions for the strong couplings of $(D, D, A)$ and $(D^{*}, D, A)$ vertices.  } \label{couplingsf1}
\begin{tabular}{|c||c|c|c|c|c|c|}
\hline
&&&&&&\\
$(D, D, A)$ &$(D^{_{-}}, \bar{D}^{_0}, a_{1}^{_-})$&$(D^{_-}, \bar{D}^{_0}, b_{1}^{_-})$&$(D_{s}^{_-}, \bar{D}^{_{0}}, K_{1A}^{_-})$&
$(D_{s}^{_-}, \bar{D}^{_{0}}, K_{1B}^{_-})$&$(D_{s}^{_+}, {D}^{_+}, K_{1A}^{_0})$&$(D_{s}^{_+}, {D}^{_+}, K_{1B}^{_0})$\\
&&&&&&\\
\hline
&&&&&&\\
$g_{_{D\,D\,A}}\,$&$0.32\pm 0.04$&$0.37\pm 0.04$&$0.89\pm 0.26$&$0.80\pm 0.21$&$0.92\pm 0.28$&$0.86\pm 0.17$\\
&&&&&&\\
\hline
\hline
&&&&&&\\
$(D^{*}, D, A)$ &$(D^{_{*-}}, \bar{D}^{_0}, a_{1}^{_-})$&$(D^{_*-}, \bar{D}^{_0}, b_{1}^{_-})$&$(D_{s}^{_*-}, \bar{D}^{_{0}},
K_{1A}^{_-})$&
$(D_{s}^{_*-}, \bar{D}^{_{0}}, K_{1B}^{_-})$&$(D_{s}^{_*+}, {D}^{_+}, K_{1A}^{_0})$&$(D_{s}^{_*+}, {D}^{_+}, K_{1B}^{_0})$\\
&&&&&&\\
\hline
&&&&&&\\
$g_{_{D^{*}\,D\,A}}\,(\rm{GeV}^{-1})$&$1.94\pm 0.63$&$2.08\pm 0.52$&$2.27\pm 0.42$&$2.06\pm 0.35$&$2.47\pm 0.64$&$2.12\pm 0.36$\\
&&&&&&\\
\hline
\end{tabular}
\end{table}
\begin{table}[th]
\caption{Couplings for the  $(D, D, V)$, $(D^{*}, D, V)$,
$(D^{*}, D^{*}\,V)$ and $(D_{1}, D_{1}, P)$ vertices.} \label{couplingsf2}
\begin{tabular}{|c||c|c|c|}
\hline
&&&\\
$(D, D, V)$ &$(D^{_-}, \bar{D}^{_0}, \rho^{_-})$&$(D_{s}^{_-}, \bar{D}^{_0}, K^{_{*-}})$&$({D}^{_0}, \bar{D}^{_0}, \psi)$
\\
&&&\\
\hline
&&&\\
$g_{_{D\,D\,V}}\,$&$1.02\pm 0.16$&$0.80\pm 0.06$&$3.03\pm 0.47$\\
&&&\\
\hline
\hline
&&&\\
$(D^{*}, D, V)$ &$(D^{_{*-}}, \bar{D}^{_0}, \rho^{_-})$&$(D_{s}^{_{*-}},
\bar{D}^{_0}, K^{_{*-}})$&$({D}^{_*0}, \bar{D}^{_0}, \psi$)\\
&&&\\
\hline
&&&\\
$g_{_{D^{*}\,D\,V}}\,(\rm{GeV}^{-1})$&$1.29\pm 0.24$&$1.06\pm 0.10$&$5.02\pm 0.66$\\
&&&\\
\hline
\hline
&&&\\
$(D^{*}, D^{*}, V)$ &$(D^{_{*-}}, \bar{D}^{_{*0}}, \rho^{_-})$&$(D_{s}^{_{*-}}, \bar{D}^{_{*0}}, K^{_{*-}})$&
$({D}^{_{*0}}, \bar{D}^{_{*0}}, \psi)$\\
&&&\\
\hline
&&&\\
$g_{_{D^{*}\,D^{*}\,V}}$&$2.22\pm 0.27$&$1.78\pm 0.21$&$5.32\pm 0.70$\\
&&&\\
\hline
\hline
&&&\\
$(D_{1}, D_{1}, P)$ &$(D_{1}^{_-}, \bar{D}_{1}^{_0}, \pi^{_-})$&$(D_{s1}^{_-}, \bar{D}_{1}^{_0}, K^{_{-}})$
&$({D}_{1}^{_0}, \bar{D}_{1}^{_0}, \eta_{c})$\\
&&&\\
\hline
&&&\\
$g_{_{D_{1}\,D_{1}\,P}}\,(\rm{GeV}^{-1})$&$0.52\pm 0.11$&$0.83\pm 0.21$&$1.35\pm 0.29$\\
&&&\\
\hline
\end{tabular}
\end{table}
To evaluate strong couplings for $A=K_{1}(1270), K_{1}(1400)$, the following relations are used:
\begin{eqnarray}
g_{D_{s}^{_-}\, \bar{D}^{_{0}}\, K_1(1270)^{_-}}&=&g_{D_{s}^{_-}\, \bar{D}^{_{0}}\, K_{1A}^{_-}}\,\sin\theta_K
+g_{D_{s}^{_-}\, \bar{D}^{_{0}}\, K_{1B}^{_-}}\,\cos\theta_K\,,\label{eq.k11}\\
g_{D_{s}^{_-}\, \bar{D}^{_{0}}\, K_1(1400)^{_-}}&=&g_{D_{s}^{_-}\, \bar{D}^{_{0}}\, K_{1A}^{_-}}\,\cos\theta_K
-g_{D_{s}^{_-}\, \bar{D}^{_{0}}\, K_{1B}^{_-}}\,\sin\theta_K\,,\label{eq.k111}\\
g_{D_{s}^{_{*-}}\, \bar{D}^{_{0}}\, K_1(1270)^{_-}}&=&r_{1A}\,g_{D_{s}^{_{*-}}\, \bar{D}^{_{0}}\, K_{1A}^{_-}}\,\sin\theta_K
+r_{1B}\,g_{D_{s}^{_{*-}}\, \bar{D}^{_{0}}\, K_{1B}^{_-}}\,\cos\theta_K\,,\label{eq.k1111}\\
g_{D_{s}^{_{*-}}\, \bar{D}^{_{0}}\, K_1(1400)^{_-}}&=&r_{2A}\,g_{D_{s}^{_{*-}}\, \bar{D}^{_{0}}\, K_{1A}^{_-}}\,\cos\theta_K
-r_{2B}\,g_{D_{s}^{_{*-}}\, \bar{D}^{_{0}}\, K_{1B}^{_-}}\,\sin\theta_K\,,\label{eq.k11111}
\end{eqnarray}
where
\begin{eqnarray}\label{eq.k12}
r_{1A}=\frac{m^{2}_{D_{s}^{_{*-}}}+m^{2}_{D_{0}}-m^{2}_{K_{1A}^{_-}}}
{m^{2}_{D_{s}^{_{*-}}}+m^{2}_{D_{0}}-m^{2}_{K_{1(1270)}^{_-}}},\,~~~~
r_{1B}=\frac{m^{2}_{D_{s}^{_{*-}}}+m^{2}_{D_{0}}-m^{2}_{K_{1B}^{_-}}}
{m^{2}_{D_{s}^{_{*-}}}+m^{2}_{D_{0}}-m^{2}_{K_{1(1270)}^{_-}}},\,\nonumber\\
r_{2A}=\frac{m^{2}_{D_{s}^{_{*-}}}+m^{2}_{D_{0}}-m^{2}_{K_{1A}^{_-}}}
{m^{2}_{D_{s}^{_{*-}}}+m^{2}_{D_{0}}-m^{2}_{K_{1(1400)}^{_-}}},\,~~~~
r_{2B}=\frac{m^{2}_{D_{s}^{_{*-}}}+m^{2}_{D_{0}}-m^{2}_{K_{1B}^{_-}}}
{m^{2}_{D_{s}^{_{*-}}}+m^{2}_{D_{0}}-m^{2}_{K_{1(1400)}^{_-}}}.
\end{eqnarray}
The $\theta_{K}$ dependence of the  strong coupling constants $g_{D_s^{-} D K_1}$ and $g_{D_s^{*-} D K_1}$
for $K_1(1270)$ and $K_1(1400)$ are displaced in Fig. \ref{fandgk1} with solid and dash lines, respectively.
The uncertainty regions are also displayed  in this figure.

Charm meson couplings to the vector, axial vector and the pseudoscalar mesons are evaluated
via different approaches. Table \ref{comcouplings}, shows the values
of the strong couplings  calculated via LCSR
\cite{momeni2020, Wanglscr2007} and  3PSR \cite{Bracco2010, Kim2001} as well as our predictions.

\begin{table}[!th]
\caption{The charm meson strong couplings in various theoretical approaches. Here, $g_{D^{_{*}}\,D\,A}$, $g_{D^{_{*}}\,D\,V}$
and $g_{D_{1}\,D_{1}\,P}$  are in the unit $\rm{GeV}^{-1}$.  } \label{comcouplings}
\begin{tabular}{|c||c|c|c||c|c|}
\hline
Coupling constant &LCSR \cite{momeni2020}&This work&Coupling constant &LCSR \cite{momeni2020}&This work\\
\hline
&&&&&\\
$g_{D^{_{-}}\,\bar{D}^{_0}\,a_{1}^{_-}}$&$0.38\pm 0.07$&$0.32\pm 0.04$&
$g_{D^{_{*-}}\,\bar{D}^{_0}\,a_{1}^{_-}}$&$1.03\pm 0.25$&$1.94\pm 0.63$\\
&&&&&\\
\hline
&&&&&\\
$g_{D^{_{-}}\,\bar{D}^{_0}\,b_{1}^{_-}}$&$1.64\pm 0.15$&$0.37\pm 0.04$&
$g_{D^{_{*-}}\,\bar{D}^{_0}\,b_{1}^{_-}}$&$1.90\pm 0.72$&$2.08\pm 0.52$\\
&&&&&\\
\hline
&&&&&\\
$g_{D_{s}^{_-}\,\bar{D}^{_{0}}\,K_{1A}^{_-}}$&$1.17\pm 0.49$&$0.89\pm 0.26$&
$g_{D_{s}^{_{*-}}\,\bar{D}^{_{0}}\,K_{1A}^{_-}}$&$1.36\pm 0.78$&$2.27\pm 0.42$\\
&&&&&\\
\hline
&&&&&\\
$g_{D_{s}^{_-}\,\bar{D}^{_{0}}\,K_{1B}^{_-}}$&$1.51 \pm 0.11$&
$0.80\pm 0.21$&$g_{D_{s}^{_{*-}}\,\bar{D}^{_{0}}\,K_{1B}^{_-}}$
&$2.48\pm 0.78$&$2.06\pm 0.35$\\
&&&&&\\
\hline
\hline
Coupling constant &LCSR \cite{Wanglscr2007}&This work&Coupling constant &LCSR \cite{Wanglscr2007}&This work\\
\hline
&&&&&\\
$g_{D^{_-}\,\bar{D}^{_0}\,\rho^{_-}}$&$1.31\pm 0.29$&$1.02\pm 0.16$&$g_{D^{_{*-}}\,\bar{D}^{_0}\,\rho^{_-}}$&$0.89\pm 0.15$&
$1.29\pm0.24$\\
&&&&&\\
\hline
&&&&&\\
$g_{D_{s}^{_-}, \bar{D}^{_0}, K^{_{*-}}}$&$1.61\pm 0.32$&$0.80\pm 0.06$&$g_{D_{s}^{_{*-}}, \bar{D}^{_0}, K^{_{*-}}}$
&$1.01\pm 0.20$&$1.06\pm 0.10$\\
&&&&&\\
\hline
\hline
Coupling constant &3PSR \cite{Bracco2010}&This work&Coupling constant &3PSR \cite{Bracco2010, Kim2001}&This work\\
\hline
&&&&&\\
$g_{{D}^{_0}, \bar{D}^{_0}, \psi}$&$5.80\pm 0.90$&$3.03\pm 0.47$&$g_{{D}^{_{*0}}, \bar{D}^{_0}, \psi}$&
$4.00\pm 0.60$&$5.02\pm 0.66$\\
&&&&&\\
\hline
&&&&&\\
$g_{{D}^{_{*0}}, \bar{D}^{_{*0}}, \psi}$&$6.20\pm 0.90$&$5.32\pm 0.70$&$g_{D_{1}^{_-}, \bar{D}_{1}^{_0}, \pi^{_-}}$
&$0.17\pm 0.04$&$0.52\pm 0.11$\\
&&&&&\\
\hline
\hline
\end{tabular}
\end{table}

\begin{figure}
\includegraphics[width=5.7cm,height=6cm]{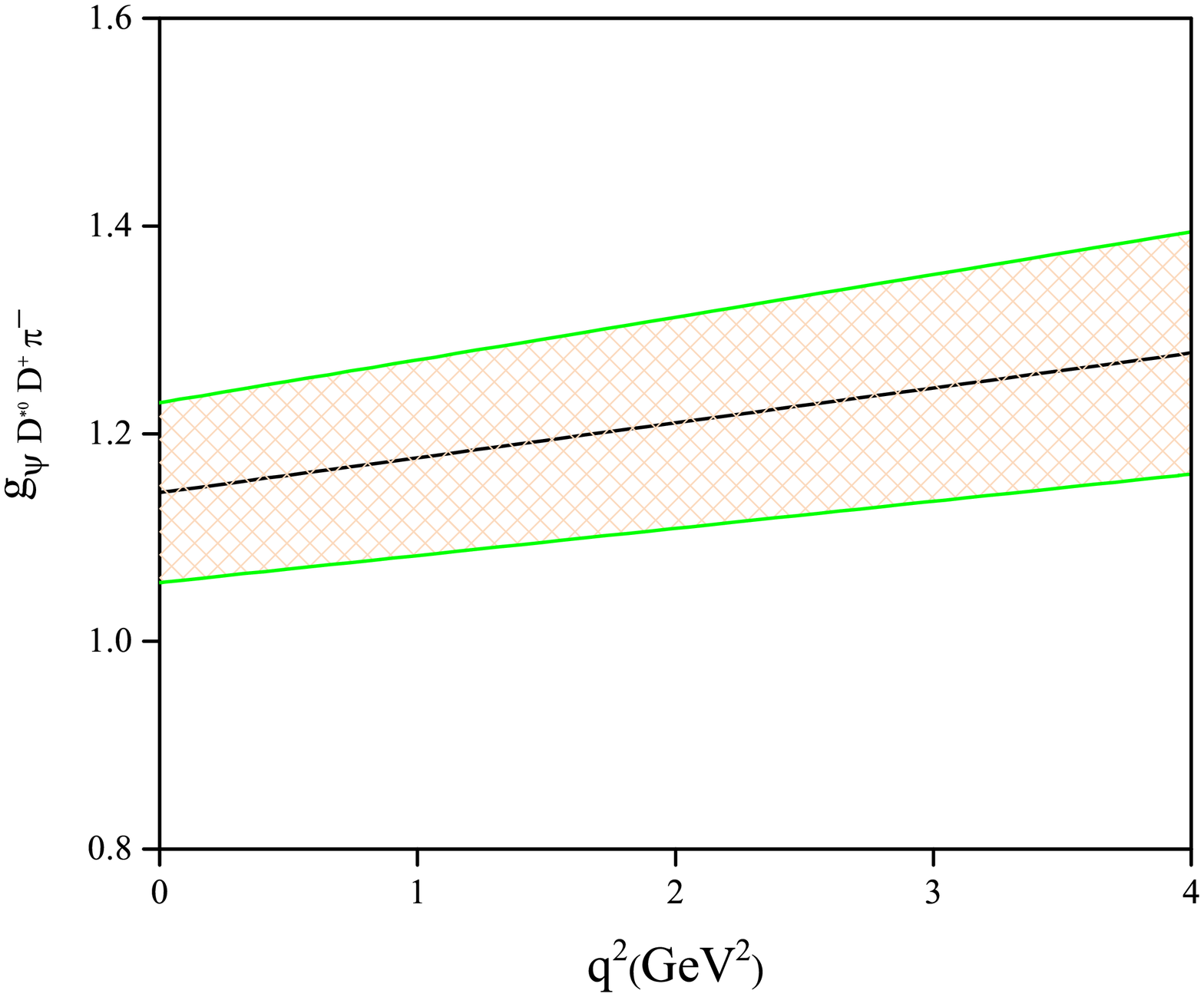}
\includegraphics[width=5.7cm,height=6cm]{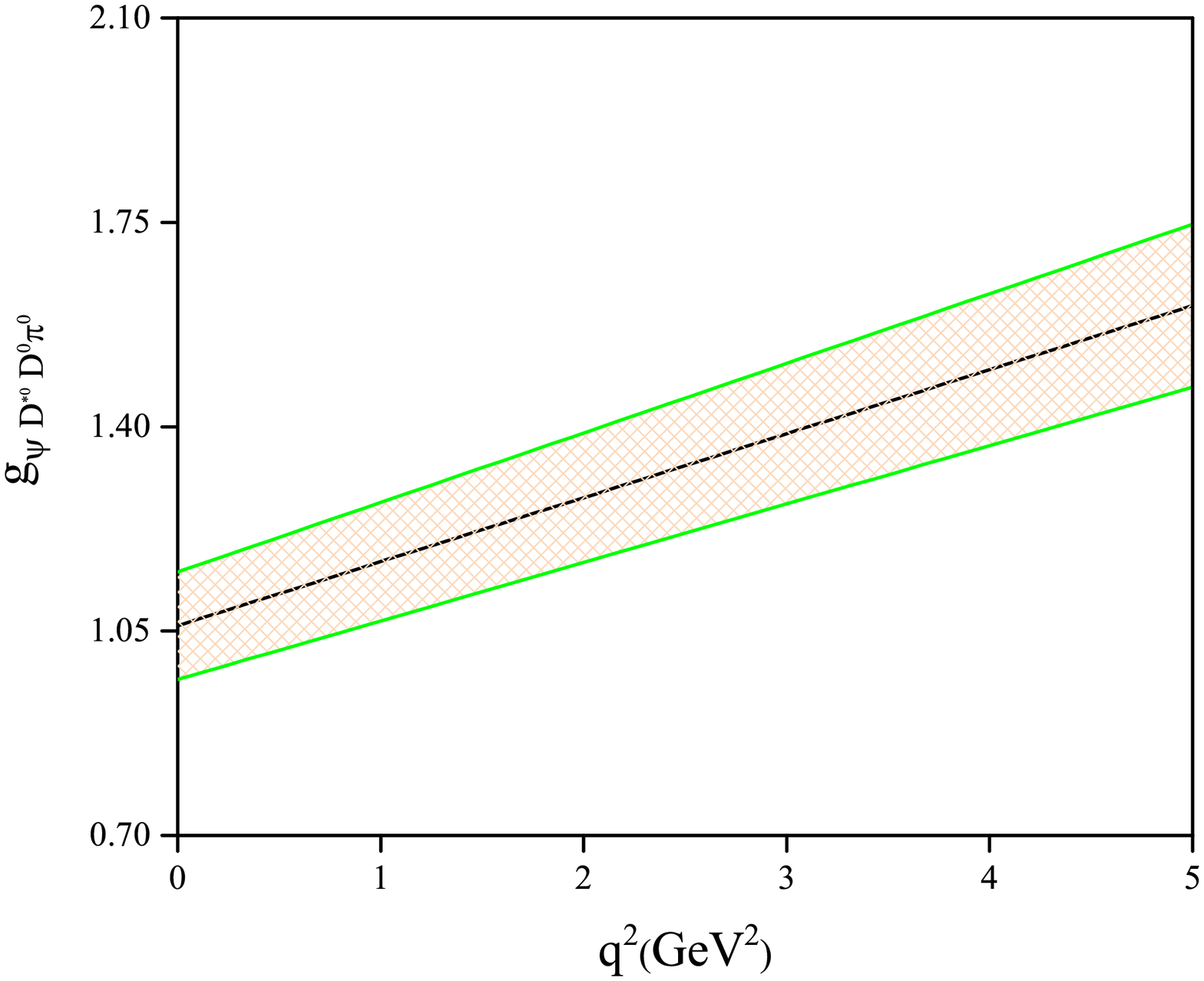}
\includegraphics[width=5.7cm,height=6cm]{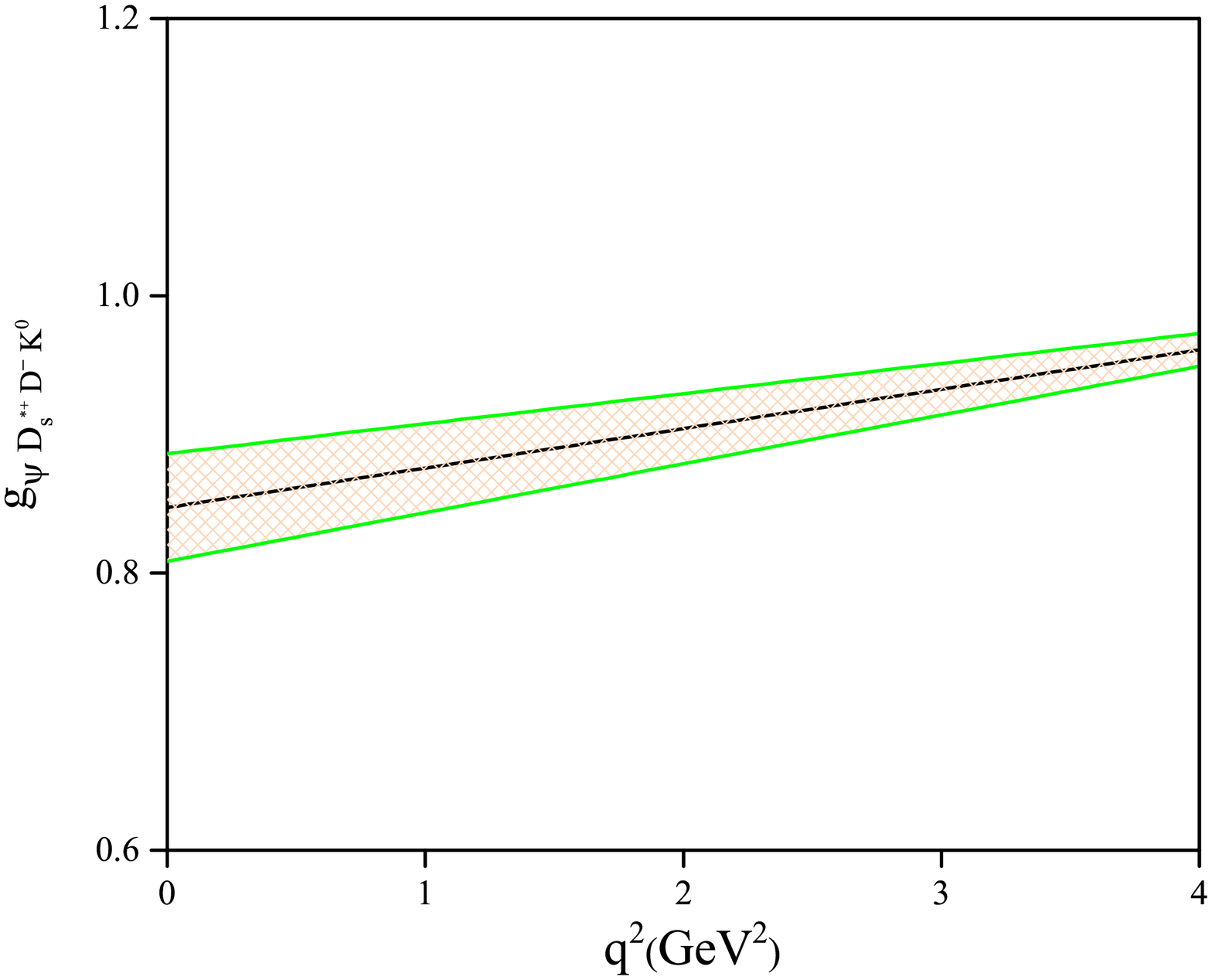}
\caption{The strong couplings of   $(\psi, D^{_{*0}}, D^{_{+}}, \pi^{_-})$, $(\psi, D^{_{*0}}, \bar{D}^{_0}, \pi^{_0})$ and
 $(\psi, D_{s}^{_{*+}}, D^{_{-}}, K^{_{0}})$ as well as their uncertainly regions on $q^2$.} \label{fourplots}
\end{figure}

Now, we consider the strong couplings for quadratic vertices. The values of $g_{\psi\,D^{(*)}\,D\,P}$
and $g_{\psi\,D^{(*)}\,D\,A}$ are listed in  Tables \ref{fourparticle1} and \ref{fourparticle2}. The reported values of
$g_{_{\psi\,D^{*}\,D\,P}}$ are at $q^2=0$.
The strong couplings
 $g_{\psi\, D^{_{*0}}\, D^{_{+}}\, \pi^{_-}}$, $g_{\psi\, D^{_{*0}}\, \bar{D}^{_0}\, \pi^{_0}}$ and
 $g_{\psi\, D_{s}^{_{*+}}\, D^{_{-}}\, K^{_{0}}}$ are plotted as functions of $q^2$ in Fig. \ref{fourplots}.
 The values of $q^2_{\rm{max}}$ are $(m_{{D}^{_+}}+m_{\pi^{_-}})^2$, $(m_{\bar{D}^{_0}}+m_{\pi^{_0}})^2$
 and $(m_{D^{_-}}+m_{K^{_{0}}})^2$ for $(\psi, D^{_{*0}}, D^{_{+}}, \pi^{_-})$,
 $(\psi, D^{_{*0}}, \bar{D}^{_0}, \pi^{_0})$ and
 $(\psi, D_{s}^{_{*+}}, D^{_{-}}, K^{_{0}})$ vertices, respectively.

\begin{figure}
\includegraphics[width=7.5cm,height=6.5cm]{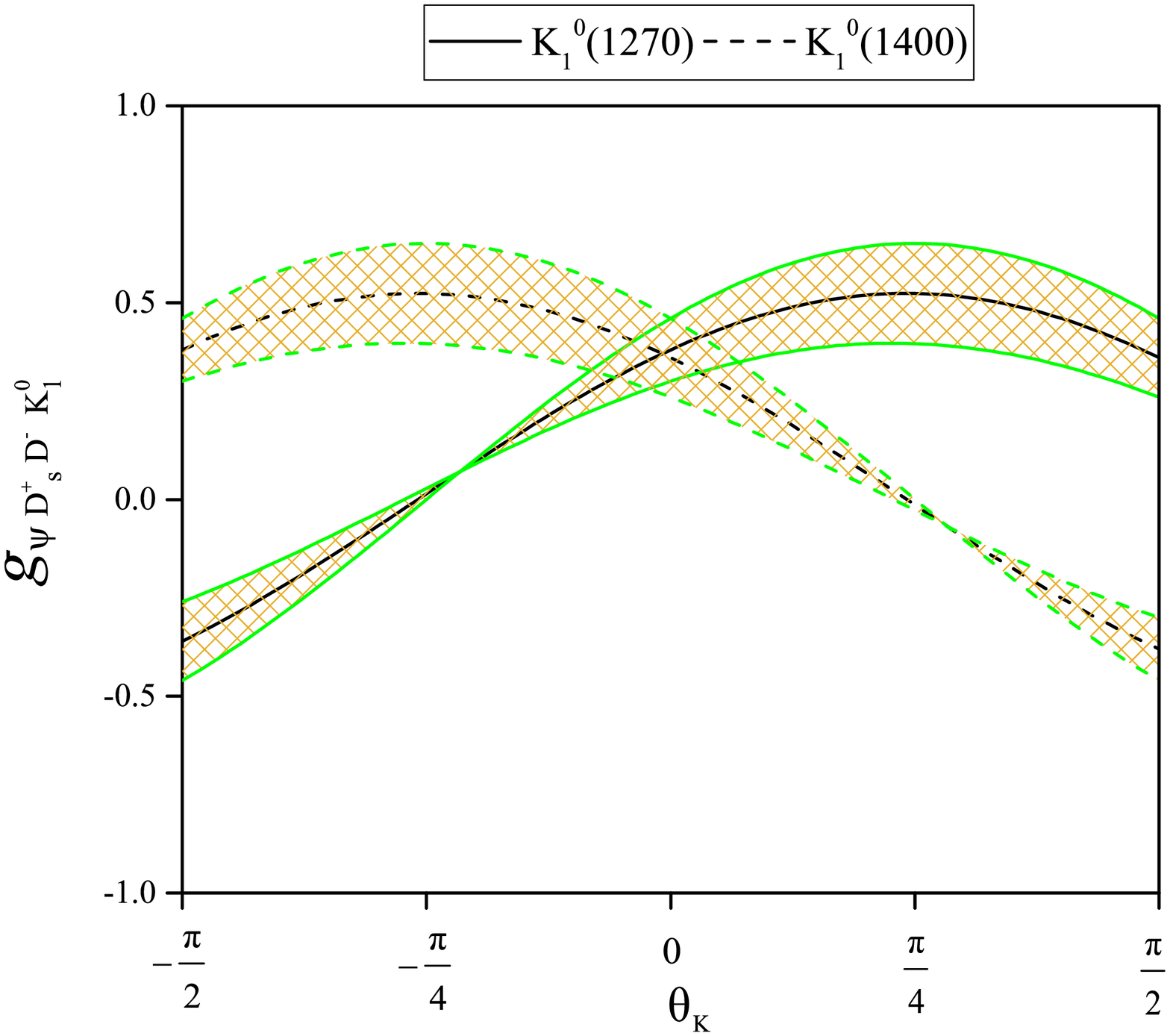}
\includegraphics[width=7.5cm,height=6.5cm]{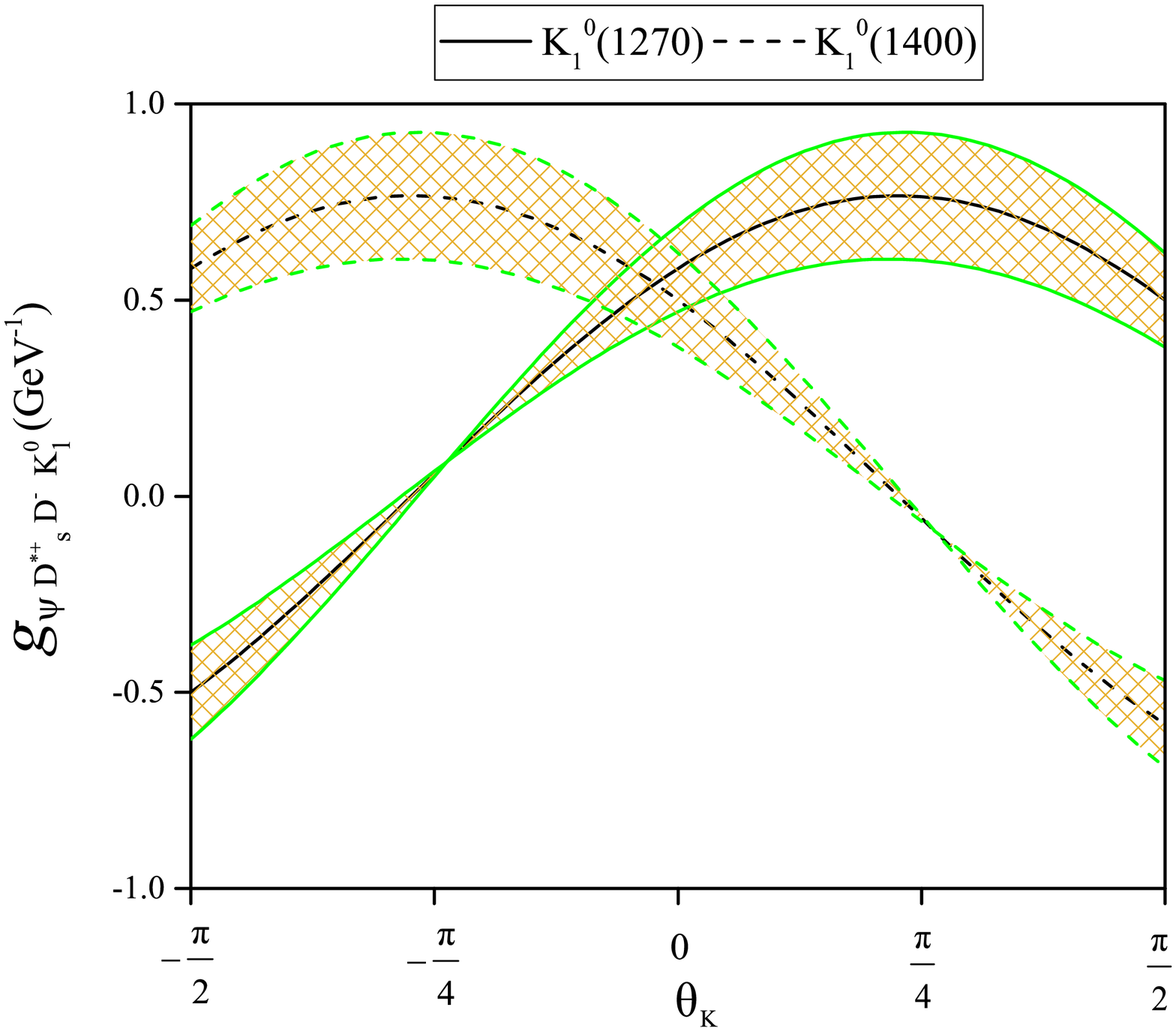}
\caption{The $\theta_{K}$ dependence of the  strong coupling constants $g_{\psi D_s^{+} D^{-} K_1^{0}}$ and $g_{\psi D_s^{*-} D^{-} K_1^{0}}$
for $K_{1}=K_1(1270),  K_1(1400)$. } \label{fourk1}
\end{figure}

\begin{table}[!th]
\caption{Our predictions for the couplings of $(\psi, D^{_{0(*0)}}, D^{_{+}}, \pi^{_-}) $,
 $(\psi, D^{_{0(*0)}}, \bar{D}^{_0}, \pi^{_0})$
and $(\psi, D_{s}^{_{+(*+)}}, D^{_{-}}, K^{_{0}})$  vertices.
 The values of $(\psi, D^{*}, D, P)$ couplings are reported at $q^2=0$. } \label{fourparticle1}
\begin{tabular}{|c||c|c|c|}
\hline
&&&\\
$( D, D, P)$ &$(D^{_0}, D^{_{+}}, \pi^{_-})$&$(D^{_0}, \bar{D}^{_0}, \pi^{_0})$
&$(D_{s}^{_{+}}, D^{_{-}}, K^{_{0}})$\\
&&&\\
\hline
&&&\\
$g_{_{\psi\,D\,D\,P}}\,~(\rm{GeV}^{-1})$&$1.28\pm 0.50$&$2.07\pm 0.85$&$0.49\pm 0.13$\\
&&&\\
\hline
&&&\\
$( D^{*}, D, P)$ &$(D^{_{*0}}, D^{_{+}}, \pi^{_-})$&$(D^{_{*0}}, \bar{D}^{_0}, \pi^{_0})$
&$(D_{s}^{_{*+}}, D^{_{-}}, K^{_{0}})$\\
&&&\\
\hline
&&&\\
$g_{_{\psi\,D^{*}\,D\,P}}\,\,(q^2=0)$&$1.14\pm 0.08$&$1.05\pm 0.10$&$0.84\pm 0.04$\\
&&&\\
\hline
\end{tabular}
\end{table}
\begin{table}[!th]
\caption{Couplings for the  $(D, D, V)$, $(D^{*}, D, V)$,
$(D^{*}, D^{*}\,V)$ and $(D_{1}, D_{1}, P)$ vertices.} \label{fourparticle2}
\begin{tabular}{|c||c|c|c|c|}
\hline
&&&&\\
$( D, D, A)$ &$(D^{_0}, D^{_{+}}, a_{1}^{_-})$&$(D^{_0}, D^{_{+}}, b_{1}^{-})$
&$(D_{s}^{_{+}}, D^{_{-}}, K_{1A}^{_{0}})$&$(D_{s}^{_{+}}, D^{_{-}}, K_{1B}^{_{0}})$\\
&&&&\\
\hline
&&&&\\
$g_{_{\psi\,D\,D\,A}}\,$&$1.27\pm 0.03$&$1.30\pm 0.05$&$0.36\pm 0.10$&$0.38\pm 0.08$\\
&&&&\\
\hline
&&&&\\
$( D^{*}, D, A)$ &$(D^{_{*0}}, D^{_{+}}, a_{1}^{_-})$&$(D^{_{*0}}, D^{_{+}}, b_{1}^{-})$
&$(D_{s}^{_{*+}}, D^{_{-}}, K_{1A}^{_{0}})$&$(D_{s}^{_{*+}}, D^{_{-}}, K_{1B}^{_{0}})$\\
&&&&\\
\hline
&&&&\\
$g_{_{\psi\,D^{*}\,D\,A}}\,\,~(\rm{GeV}^{-1})$&$0.12\pm 0.02$&$0.14\pm 0.02$&$0.50\pm 0.12$&$0.58\pm 0.11$\\
&&&&\\
\hline
\end{tabular}
\end{table}
 To evaluate the couplings of $(\psi,  D_{s}^{_{+}},  D^{_{-}},  K_{1}^{_{0}}(1270))$,
$(\psi, D_{s}^{_{+}},  D^{_{-}},  K_{1}^{_{0}}(1400))$,  $(\psi, D_{s}^{_{*+}}, D^{_{-}}, K_{1}^{_{0}}(1270))$
 and $(\psi, D_{s}^{_{*+}}, D^{_{-}}, K_{1}^{_{0}}(1400))$ vertices,
 we use the relations  similar to those   used in Eqs. (\ref{eq.k11}) and (\ref{eq.k111}).
 These couplings and their uncertainly regions
 are plotted as functions of the mixing angle $\theta_{K}$ in Fig \ref{fourk1}. Our numeric analyze show that the main sources of
uncertainties in the four particles vertices are $m_{c}$ and $\sigma_{c}$.

In summary  in this paper  the two
flavor hard-wall holographic
model introduced in \cite{Erlich2005} is extended  to four
flavors. Our model consists of  nine parameters including the hard wall position $z_{0}$, quark masses $m_{q}$  and
quark condensates $\sigma_{q}$ with $q=(u, d, s, c)$. These
parameters are fitted to the  experimental masses  of
$\rho^{_0}$, $\rho^{_-}$, $a_{1}^{_{-}}$,
$\pi^{_0}$,  $\pi^{_-}$
$K^{_-}$,  $K^{_{*-}}$, $D^{_{-}}$ and $D^{_{*-}}$ mesons.
The masses  and decay constants of some pseudoscalar, vector and axial vector mesons are
evaluated  using our model and a comparison is made between our predictions
and the experimental data for these observables.
After analyzing the wave functions, the strong couplings of
$(D^{(*)}, D, A)$, $(D^{(*)}, D^{(*)}, V)$, $(D_{1}, D_{1}, P)$,
 $(\psi, D^{(*)}, D, P)$ and $(\psi, D^{(*)}, D, A)$ are
analyzed. For $A=(K_{1}(1270), K_{1}(1400))$ the strong couplings are plotted as functions
of the mixing angle $\theta_{K}$. Moreover,  for three mesons vertices  a comparison is also made between our
predictions and  estimations  made by  other theoretical approaches.

\clearpage
\appendix
\section{ values for  $g^{abc}$, $h^{abc}$, $l^{abc}$, $k^{abc}$,
 $h^{abcd}$, $k^{abcd}$ and $l^{abcd}$}\label{appA:factors}
In this appendix, we present the nonzero values for $g^{abc}$,
$h^{abc}$, $l^{abc}$, $k^{abcd}$, $h^{abcd}$,  $k^{abcd}$ and $l^{abcd}$.
The values results of the  factors appeared in 3-point functions which are used in numerical analyze
are given in Table \ref{Tafghlk}.

\begin{table}[th]
\caption{The values of $g^{abc}$,
$h^{abc}$, $l^{abc}$ and  $k^{abc}$ which are used in numerical analyze. } \label{Tafghlk}
\begin{ruledtabular}
\begin{tabular}{|ccccc|}
&&&&\\
$(a, b, c)$ &$g^{abc}$&$h^{abc}$&$l^{abc}$&$k^{abc}$\\
&&&&\\
\hline
&&&&\\
$(2, 9, 11)$ &$\frac{1}{2}(v_{u}+v_{d})(v_{u}+v_{c}) $&
$\frac{1}{2}(v_{u}-v_{d})(v_{u}+v_{c})$&$\frac{1}{2}(v_{d}+v_{c})(v_{u}+v_{c})$&$\frac{1}{2}(v_{d}-v_{u})(v_{d}+v_{c}) $\\
&&&&\\
$(4, 9, 14)$&$\frac{1}{2}(v_{u}+v_{s})(v_{u}+v_{c})$&
$\frac{1}{2}(v_{u}-v_{s})(v_{u}+v_{c})$&$\frac{1}{2}(v_{s}+v_{c})(v_{u}+v_{c})$&$\frac{1}{2}(v_{u}-v_{s})(v_{u}+v_{c})$\\
&&&&\\
$(6, 11, 14)$&$-\frac{1}{2}(v_{d}+v_{s})(v_{s}+v_{c})$&
$-\frac{1}{2}(v_{s}-v_{d})(v_{d}+v_{c})$&$\frac{1}{2}(v_{d}+v_{c})(v_{s}+v_{c})$&$-$\\
&&&&\\
$(9, 10, 15)$ &$\frac{\sqrt{6}}{6}(v_{u}+v_{c})(v_{u}+3\,v_{c})$&
$\frac{\sqrt{6}}{6}(v_{u}-v_{c})(v_{u}-3\,v_{c})$&$-$&$\frac{\sqrt{6}}{6}(v_{d}-v_{c})(v_{d}+3\,v_{c})$\\
&&&&\\
\end{tabular}
\end{ruledtabular}
\end{table}

\begin{table}[th]
\caption{The values of  $g^{abcd}$,
$h^{abcd}$, $l^{abcd}$ and  $k^{abcd}$ which are used in numerical analyze. } \label{Tafghlk}
\begin{ruledtabular}
\begin{tabular}{|ccccc|}
&&&&\\
$(a, b, c, d)$&$-i\,g^{abcd}$&$-i\,h^{abcd}$&$-i\,l^{abcd}$&$-i\,k^{abcd}$\\
&&&&\\
\hline
&&&&\\
$(9, 15, 2, 11)$&
$\frac{\sqrt{6}}{6}(v_{u}+v_{c})(v_{d}+v_{c})$&$-\frac{\sqrt{6}}{12}(v_{u}-v_{c})(v_{d}+v_{c})$&
$-\frac{\sqrt{6}}{12}(v_{d}+v_{c})(v_{u}+3\,v_{c})$&$-\frac{\sqrt{6}}{12}(v_{u}+v_{d})(v_{u}+v_{c})$\\
&&&&\\
\hline
&&&&\\
$(9, 15, 3, 10)$&
$\frac{\sqrt{6}}{6}(v_{u}+v_{c})^2$&$-\frac{\sqrt{6}}{12}(v_{u}^2-v_{c}^2)$&$
-\frac{\sqrt{6}}{12}(v_{u}+v_{c})(v_{u}+3\,v_{c})$&$-\frac{\sqrt{6}}{6}v_{u}\,(v_{u}+v_{c})$\\
&&&&\\
\hline
&&&&\\
$(13, 15, 7, 11)$&
-$\frac{\sqrt{6}}{6}(v_{s}+v_{c})(v_{d}+v_{c})$&$
\frac{\sqrt{6}}{12}(v_{d}+v_{c})(v_{s}-v_{c})$&$\frac{\sqrt{6}}{12}(v_{d}+v_{c})(v_{s}+3\,v_{c})$&
$-\frac{\sqrt{6}}{12}(v_{d}+v_{c})(v_{s}+v_{c})$\\
&&&&\\
\end{tabular}
\end{ruledtabular}
\end{table}

\end{document}